\documentclass[11pt]{article}
\usepackage{amsmath}
\usepackage{amsfonts}
\usepackage{amssymb}
\usepackage{graphicx}
\usepackage{booktabs}
\usepackage{lineno}
\usepackage{setspace}
\usepackage{comment}
\usepackage{caption}
\usepackage{subcaption}
\usepackage{tikz}
\usepackage{hyperref}
\usetikzlibrary{positioning, arrows.meta}

\tikzset{
    stepbox/.style = {rectangle, rounded corners, draw=black, minimum width=10cm, font=\small, align=center, inner sep=8pt},
    data/.style   = {stepbox},
    mesh/.style   = {stepbox},
    uncert/.style = {stepbox},
    cvfold/.style = {stepbox},
    eval/.style   = {stepbox},
    pca/.style    = {stepbox},
    sensA/.style  = {stepbox},
    grey/.style   = {stepbox},
    arr/.style    = {-{Stealth[scale=1.0]}, thick, gray},
}

\title{Which modelling choices shape your interpolation map?\\
An Optimal Transport sensitivity analysis for groundwater sulfate concentration}
\author{Corinna Perchtold\thanks{Bureau de Recherches Géologiques et Minières (BRGM) -- HOUSES project (ANR-22-CE56-0006)} \and Jérémy Rohmer \and Augustin Thomas \and Julie Lions}

\begin{document}
\maketitle







\begin{abstract}
Spatial prediction maps are crucial for environmental management yet highly sensitive to subjective modelling choices such as, e.g., covariates, spatial dependence structures, or mesh resolution.
We develop a workflow that attributes the resulting map uncertainty to these choices and is applicable to any spatial model, since it operates only on an ensemble of maps produced by propagating the choices through spatial prediction.
This ensemble is then filtered to retain only models with acceptable  predictive performance, reduced in dimension by principal component (PC) analysis, and the sensitivity of the resulting scores to each choice is quantified by 
global sensitivity analysis based on optimal transport (OT) metrics.

 We demonstrate the framework with Bayesian geostatistics  
 applied to  the spatial mapping of sulphate concentrations in groundwater of Paris Basin (France), which is key for assessing groundwater chemical quality and establishing the natural geochemical baseline of aquifers.
For this application, the covariates, mesh resolution, covariance models, and priors are combined into 18000 candidate configurations. These are filtered by cross-validated root mean square error (RMSE) and 95\% coverage of the prediction interval, and the retained configurations are then inferred on a high-resolution grid and reduced to its three PCs.
 The OT analysis identifies covariate selection  (under 95\% coverage) and mesh resolution (under RMSE) as the dominant sources of uncertainty to the PCs,   and a negligible one for the choice of the  covariance model. 
\end{abstract}

\section{Introduction}
Spatial prediction maps increasingly support monitoring and management decisions across the environmental sciences. 
Yet such maps are only as trustworthy as the modelling choices behind them, and those choices are rarely unique: model and covariate specification can dominate the prediction error of a spatial map (\cite{Nelson2011}), and treating a single chosen configuration as if it were certain understates the predictive uncertainty (\cite{Pilz08}).
A practitioner mapping a spatially distributed quantity must decide 
which covariance structure to adopt, which priors to place on it if a Bayesian strategy is chosen, and which covariates to include and in what functional form. Each a defensible choice leading to equally satisfactory performing spatial models. These decisions propagate into the final map, but their relative influence on its reliability is seldom quantified. In part because, to our knowledge, no  unified workflow or set of tools for doing so is available in the literature.

Our objective is to propose such a workflow that aims to quantify how strongly each choice shapes the resulting prediction map.
To this end we do not commit to one configuration but generate an ensemble of maps that the admissible choices produce, and attribute the variation across this ensemble to the individual
choices.
 We showcase the workflow on groundwater sulphate concentrations in the Paris Basin. 
Because the workflow operates on such an ensemble, it is agnostic to the inference
engine and applies equally to kriging, Bayesian geostatistical models, or machine learning methods for spatial data. 
What changes between engines is only the set of controllable factors under study.

Where exactly these choices enter becomes clear from the structure of a spatial model.
 Consider a target variable observed at $n$ georeferenced locations $\mathbf{s}_1,\dots,\mathbf{s}_n$ in a spatial domain $\mathcal{D} \subset \mathbb{R}^2$. 
Spatial models  (\cite{Cressie1991}, \cite{Banerjee2014}) are generally written as \begin{equation}\label{eq:spatial_model}
    Y(\mathbf{s})= f(\mathbf{s},\beta) +Z(\mathbf{s})
\end{equation}  with $ \mathbf{s} \in \mathcal{D}$, where $f(\cdot,\beta)$ is the conditional mean capturing the systematic variation explained by the covariates and their fixed effects $\beta$, and $Z(\cdot)$ is a spatially dependent random process describing the residual dependence through its covariance function.
 Both components carry the uncertainty we set out to investigate:  the mean requires selecting the covariates and the functional form of their effects,
while  the spatial process a covariance family and its parameters. Every admissible specification yields a different prediction map.
The difficulty lies not in fitting any single model but in systematically attributing the uncertainty of the resulting map to the modelling choices that produced it. Providing such an attribution is the objective of this paper.

To do so, we rely on the framework of global sensitivity analysis (GSA, \cite{Saltelli2008}), which we adapt to our setting in three respects.
The first challenge is dimensionality: the output is not a scalar but a prediction map over a dense grid, typically comprising several thousand pixels. 
 To address this, we follow the strategy advocated in the literature (\cite{LAMBONI2013, Marrel2010, PERRIN2021}) and apply a dimension reduction method, here principal component (PC) analysis.
 The second is the nature of the effect to be measured: a modelling choice may reshape the output distribution in any respect, not only in its variance, so a variance-based  index such as Sobol' would miss part of its influence. 
 This calls for a moment-independent measure, which quantifies the effect of an input on the entire distribution (\cite{BORGONOVO2007}). 
 The third is multivariateness: even after reduction, the output is not a single PC
score but a joint distribution of PC score vectors, so the measure must compare distributions in several dimensions rather than one. 
We therefore combine PC analysis with sensitivity indices based on optimal transport (OT), recently proposed by \cite{Borgonovo2025}. 
These indices compare the unconditional distribution of the output with its distribution conditional on fixing an input, and are particularly suited to our setting because they account by construction for the multivariate nature of the output.

While OT-based GSA has been developed for dynamic and functional outputs (\cite{Chiani2025b}), to our knowledge it has not previously been used to quantify the sensitivity of geostatistical prediction maps to modelling choices.
 We close this gap  by combining OT-based sensitivity measures with a PCA representation of the map ensemble.

Our contributions are twofold. First,  we bring OT sensitivity measures to bear on geostatistical prediction maps, revealing where and how strongly each modelling choice affects the resulting map. 
Second, we assess for the first time on a real case the joint contribution of multiple modelling choices, with an application to  groundwater sulphate concentration mapping in the Paris Basin.
We choose the SPDE-INLA approach to Bayesian geostatistical inference
(\cite{Rue2009, Lindgren2011}), which couples the stochastic partial differential equation (SPDE) representation of the spatial field, whose covariance belongs to the Matérn family, with the integrated nested Laplace approximation (INLA) for inference.
It provides full posterior uncertainty for every prediction map, and its efficient \texttt{R} implementation (\cite{RTeam2011}) and extensive supporting literature (\cite{Krainski2019, Blangiardo2015, Bakka2018}) make repeated fitting across many scenarios routine.

The remainder of this paper is organised as follows. Section~\ref{sec:method}  presents the methodological framework, a six-stage workflow that propagates modelling choices through spatial prediction and quantifies their influence through optimal transport sensitivity analysis.
Section~\ref{sec:data} describes the study area, the groundwater sulphate data set, and the environmental covariates, and specifies the spatial model and inference method used to instantiate the workflow.
 Section~\ref{sec:results} presents the results of applying the workflow to the Paris Basin case study, with particular emphasis on how our modelling choices contribute to uncertainty in the resulting spatial predictions. 
Section~\ref{sec:discussion} assesses the sensitivity of the results to the choice of performance thresholds in the filtering stage.
Finally, Section~\ref{sec:conclusion} discusses the main findings, methodological limitations, and practical implications of the proposed framework.

\section{Methodology}
\label{sec:method}
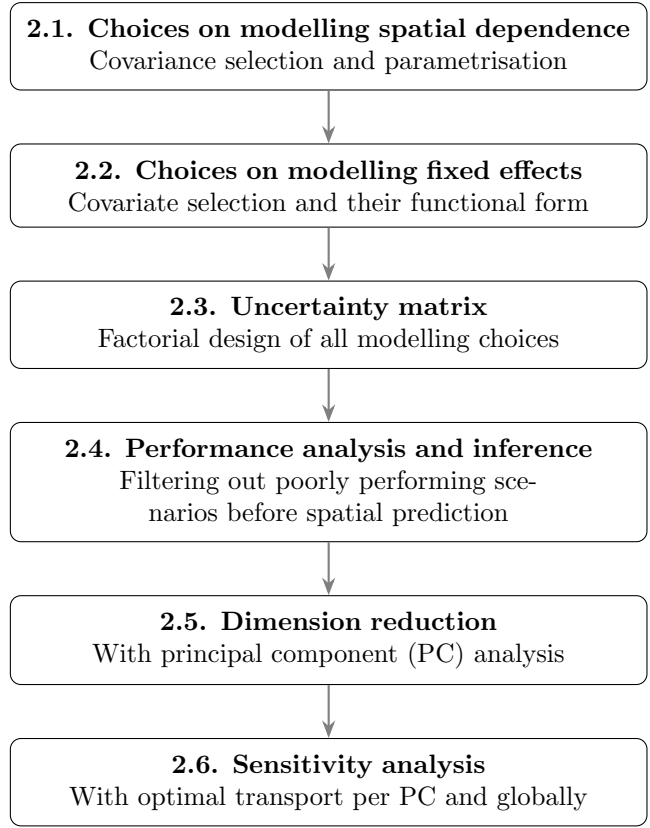
\begin{figure}[h]
\centering
\begin{tikzpicture}[
    node distance=0.7cm,
    align=center,
    stepbox/.style={rectangle, rounded corners, draw=black, text width=8cm,
                    minimum height=1.1cm, font=\small, align=center, inner sep=6pt},
    arr/.style={-{Stealth[scale=1.0]}, thick, gray}
]
\node[stepbox, data]              (s1)            {\textbf{2.1.  Choices on modelling spatial dependence}\\ Covariance selection and parametrisation};
\node[stepbox, mesh,  below=of s1] (s2)           {\textbf{2.2. Choices on modelling fixed effects} \\  Covariate selection and  their functional form};
\node[stepbox, uncert,below=of s2] (s3)           {\textbf{2.3. Uncertainty matrix} \\ Factorial design of all modelling choices};
\node[stepbox, cvfold,below=of s3] (s4)           {\textbf{2.4. Performance analysis and inference} \\ Filtering out poorly performing scenarios before spatial prediction};
\node[stepbox, pca,   below=of s4] (s5)           {\textbf{2.5. Dimension reduction} \\ With principal component (PC) analysis};
\node[stepbox, sensA, below=of s5] (s6)           {\textbf{2.6. Sensitivity analysis} \\ With optimal transport per PC and globally};

\draw[arr] (s1) -- (s2);
\draw[arr] (s2) -- (s3);
\draw[arr] (s3) -- (s4);
\draw[arr] (s4) -- (s5);
\draw[arr] (s5) -- (s6);
\end{tikzpicture}
\caption{Model-generic six-stage spatial sensitivity workflow.}
\label{fig:workflow}
\end{figure}

We introduce a six-stage workflow (Figure~\ref{fig:workflow}) for analysing the influence of modelling
inputs on the spatial prediction map of an environmental phenomenon.
We instantiate the workflow 
with a model specified by a random effect modelled through a  Gaussian random field  with a covariance function that captures the spatial dependence structure  (Subsection~\ref{sec:inla}) and  by a mean function that accounts for the effects of multiple covariates (Subsection~\ref{sec:covariates}).
The modelling choices under study are catalogued in an uncertainty matrix (Subsection~\ref{sec:matrix}) that records each controllable factor together with its admissible levels.
All combinations of these levels form an ensemble of candidate model configurations, each yielding a spatial map. 
Predictive performance criteria then screen this ensemble (Subsections~\ref{sec:CV}), discarding configurations that produce implausible maps and restricting the subsequent inference to the retained scenarios.
In order to reduce the dimension of the high-resolution spatial prediction maps we perform PCA (Subsections~\ref{sec:pca-method}) and analyses the resulting global and per-PC scores with sensitivity analysis based on optimal transport  (Subsection~\ref{sec:sa-method}).

\subsection{Choices on modelling spatial dependence}

\label{sec:inla}
The first step is to establish whether spatial dependence is present. Fitting Equation~\ref{eq:spatial_model} through an intercept-only model and inspecting the residuals for spatial structure indicates whether nearby locations tend to have
similar target values. Gaussian stochastic processes are widely used to model such geostatistical dependence (\cite{Diggle07}).
Such a process is characterised by a covariance function that specifies how the correlation between two locations decays with the distance separating them, governed by a few parameters, typically a correlation range and a marginal variance, which control how far and how strongly observations influence one another. 
The Matérn covariance family (\cite{Matern1960}) is a particularly popular choice in geostatistical modelling.
However, the choice of covariance family and its associated parameters is rarely unique, with several specifications often providing a plausible representation of the spatial dependence structure. These choices consequently constitute a source of structural uncertainty that is explicitly accounted for in the proposed workflow.

\subsection{Choices on modelling fixed effects}
\label{sec:covariates}
Alongside the spatial random effect, a structured additive framework
(\cite{Fahrmeier01}) lets the
target variable depend on covariates through fixed effects, see Equation~\ref{eq:spatial_model}.
Covariates are auxiliary variables observed at each
location, e.g., physical or geological attributes, that carry explanatory information about the target variable.
Two choices concern the mean structure. 
First, the likelihood of the target variable is chosen to match its distributional shape. 
Second, the functional form of each covariate effect is decided, either linear or non-linear. 
In our case, we allow first-order random-walk (RW1) latent effects in place of linear fixed effects. The RW1 specification treats each covariate as an ordered sequence and smooths the effect across adjacent values, capturing plateaus and other non-monotonic structure that a linear term would average out. 
The candidate likelihoods and covariate forms are evaluated jointly by two negatively oriented Bayesian criteria: the Deviance Information Criterion (DIC; \cite{Spiegelhalter2014}) and the Watanabe--Akaike Information Criterion (WAIC; \cite{Gelman2014}). The DIC evaluates model quality by balancing the posterior mean deviance, a measure of goodness-of-fit, against the effective number of parameters, which penalises model complexity and protects against overfitting in high-dimensional spatial models.
Complementary to this, the WAIC uses the full pointwise posterior predictive density and is the more robust estimator of out-of-sample accuracy, as it does not rely on a posterior point estimate. 
Then the likelihood is fixed and the covariate selection and their forms enter the uncertainty matrix  as a factor.

\subsection{The uncertainty matrix}
\label{sec:matrix}
Subsections~\ref{sec:inla}   and~\ref{sec:covariates}  each surfaced modelling decisions with no unambiguous answer. 
 Rather than committing to one configuration and reporting a single prediction map, we treat each decision as a controllable factor and construct the full factorial combination of their levels, collected into a matrix $\mathbf{U} \in \mathbb{R}^{N \times p}$ which we term the \emph{uncertainty matrix}. 
Each of the $p$ columns corresponds to one factor, i.e., one modelling decision, and takes a finite set
of levels representing the admissible alternatives for that decision.
Each of the $N$ rows is one fully specified model, or scenario, obtained by fixing every factor to one of its levels, i.e., $N$ is therefore the product of the per-factor level counts. Spanning this joint space lets the GSA 
attribute variability in the resulting prediction maps to individual factors. 

\subsection{Performance analysis and inference}
\label{sec:CV}
Varying the modelling choices across the full factorial produces many scenarios whose predictive performance is poor, and which are therefore implausible as representations of the underlying process. 
Attributing the uncertainty of the prediction map across all $N$ scenarios would let these implausible models dominate the result. 
We therefore assess the predictive performance of the $N$ scenarios 
and retain only the near-optimal subset, i.e., scenarios that achieve near-identical empirical performance yet may differ substantially in their predictions. A set known  as a Rashomon set (\cite{Breiman01}). 
Quantifying the sensitivity within
this set isolates the disagreement that persists among models that the data
cannot distinguish, which is precisely the uncertainty of practical interest.

To assess the predictive performance we adopt a   Leave-Group-Out cross-validation (CV) scheme  for spatially structured models (\cite{liu2025}).
It removes clusters of dependent observations simultaneously, forcing the model to predict from more distant data and giving a more realistic estimate of out-of-sample performance (\cite{Adin2024, liu2025}) than standard Leave-One-Out CV. The size of the removed cluster trades a genuine spatial buffer against retaining enough training data, and is chosen accordingly for the application at hand.
We 
evaluate each scenario on two complementary axes: the accuracy of the point prediction, through the RMSE 
and the calibration of the predictive uncertainty, through the coverage of the posterior predictive intervals:

\begin{itemize}

    \item \textbf{Root Mean Square Error (RMSE):} Measures the average magnitude of prediction errors, 
    i.e., 
    \begin{equation*}
        \text{RMSE} = \sqrt{\frac{1}{n}\sum_{i=1}^{n}(\hat{y}_i - y_i)^2},
    \end{equation*}
    where $y_i$ is the sulphate concentration and $\hat{y}_i$ is the posterior predictive mean at held-out station $i$ and  $n$ the number of stations. Lower values indicate better point prediction accuracy.

    \item \textbf{Prediction interval coverage:} The empirical proportion of held-out observations falling within the posterior predictive interval,
    \begin{equation*}
        \text{Coverage}_{(1-\alpha)} = 
        \frac{1}{n}\sum_{i=1}^{n} 
        \mathbf{1}(l_{\alpha,i} \leq y_i \leq u_{\alpha,i}).
    \end{equation*}
    A well-calibrated model should achieve coverage close to the nominal level: approximately 
    0.95 for the 95\% interval. 

\end{itemize}

\noindent
A scenario is retained if it ranks among the best-performing under at least one of these metrics. 
Filtering this way retains a subset of $\hat{N} < N$ scenarios, giving the reduced uncertainty matrix $\hat{\mathbf{U}} \in \mathbb{R}^{\hat{N} \times p}$: the $p$ factor columns of $\mathbf{U}$ restricted to the retained rows. 
For each of the $\hat{N}$ retained scenarios, 
we infer the target variable at every point of a high-resolution grid of $G$ locations covering the domain. 
For a mean analysis extract the predictive mean.
Collecting these over all retained scenarios yields 
the ensemble mean matrix 
$\mathbf{M} \in \mathbb{R}^{\hat{N} \times G}$.

\subsection{Dimension reduction}
\label{sec:pca-method}

We reduce the filtered map ensemble using principal component analysis (PCA), which represents the data in an orthogonal basis ordered by explained variance. Retaining only the leading components provides a compact description of scenario differences, with modelling choices reflected as shifts in the component-score distributions (\cite{Jolliffe16}).
We apply it to the matrix $\mathbf{M}$.  
Each grid point is centred and scaled to zero mean and unit variance across the $\hat{N}$ scenarios, so the decomposition acts on the correlation rather than
the covariance matrix (\cite[\S2(c)]{Jolliffe16}), preventing grid points with large absolute concentrations from dominating through scale alone.
We retain the smallest number of components $K$ whose leading PCs together explain at least $C\%$ of the total
variance,
\[
K = \min\Big\{k \;:\; \sum_{r=1}^{k} d_r \big/ \sum_{j} d_j
\;\geq\; \frac{C}{100}\Big\},
\]
where $d_r$ is the variance of the $r$-th component. The decomposition then yields two objects:
\begin{itemize}
    \item the spatial loadings (eigenvectors) $\mathbf{v}_1, \dots, \mathbf{v}_K\in \mathbb{R}^G$. Mapped onto the domain, each $\mathbf{v}_r$ is a PC map displaying one spatial pattern of cross-scenario variation. 
    \item the scores $\mathbf{B} \in \mathbb{R}^{\hat{N} \times K}$. 
Row $i$ gives the coordinates of retained scenario $i$'s map in the $K$-dimensional reduced space. A score is itself non-spatial.
\end{itemize}
The scores $\mathbf{B}$ are the input to the
sensitivity analysis. 

Additionally, we summarise the ensemble matrix $\textbf{M}$  by the coefficient of variation (CoV). With $M_{ig}$ 
the predictive mean matrix of retained scenario $i$ at grid point $g$, the CoV at grid point $g$ is computed down each column of $\mathbf{M}$,
\begin{equation*}
    \mathrm{CoV}(g) \;=\;
    \frac{\displaystyle \sqrt{\frac{1}{\hat{N}-1}\sum_{i=1}^{\hat{N}}
            \big(M_{ig} - \overline{M}_{\cdot g}\big)^2}}
         {\left|\dfrac{1}{\hat{N}}\displaystyle\sum_{i=1}^{\hat{N}} M_{ig}\right|}
    \;\times\; 100\%,
\end{equation*}
where
\begin{itemize}
    \item $g \in \{1,\dots,G\}$ indexes the $G$ grid points,
    \item $i \in \{1,\dots,\hat{N}\}$ indexes the $\hat{N}$ retained scenarios,
    \item $\overline{M}_{\cdot g} = \tfrac{1}{\hat{N}}\sum_{i=1}^{\hat{N}} M_{ig}$ is the cross-scenario mean at grid point $g$.
\end{itemize}
It is a scale-free measure of how strongly the $\hat{N}$  retained scenarios disagree at a grid point $g$ relative to the local prediction level. 
The PC maps complement the CoV by showing the spatial pattern of that disagreement rather than its pointwise magnitude.

\subsection{Sensitivity analysis with optimal transport}
\label{sec:sa-method}

\begin{figure}[h]
    \centering
\includegraphics[width=0.8\linewidth]{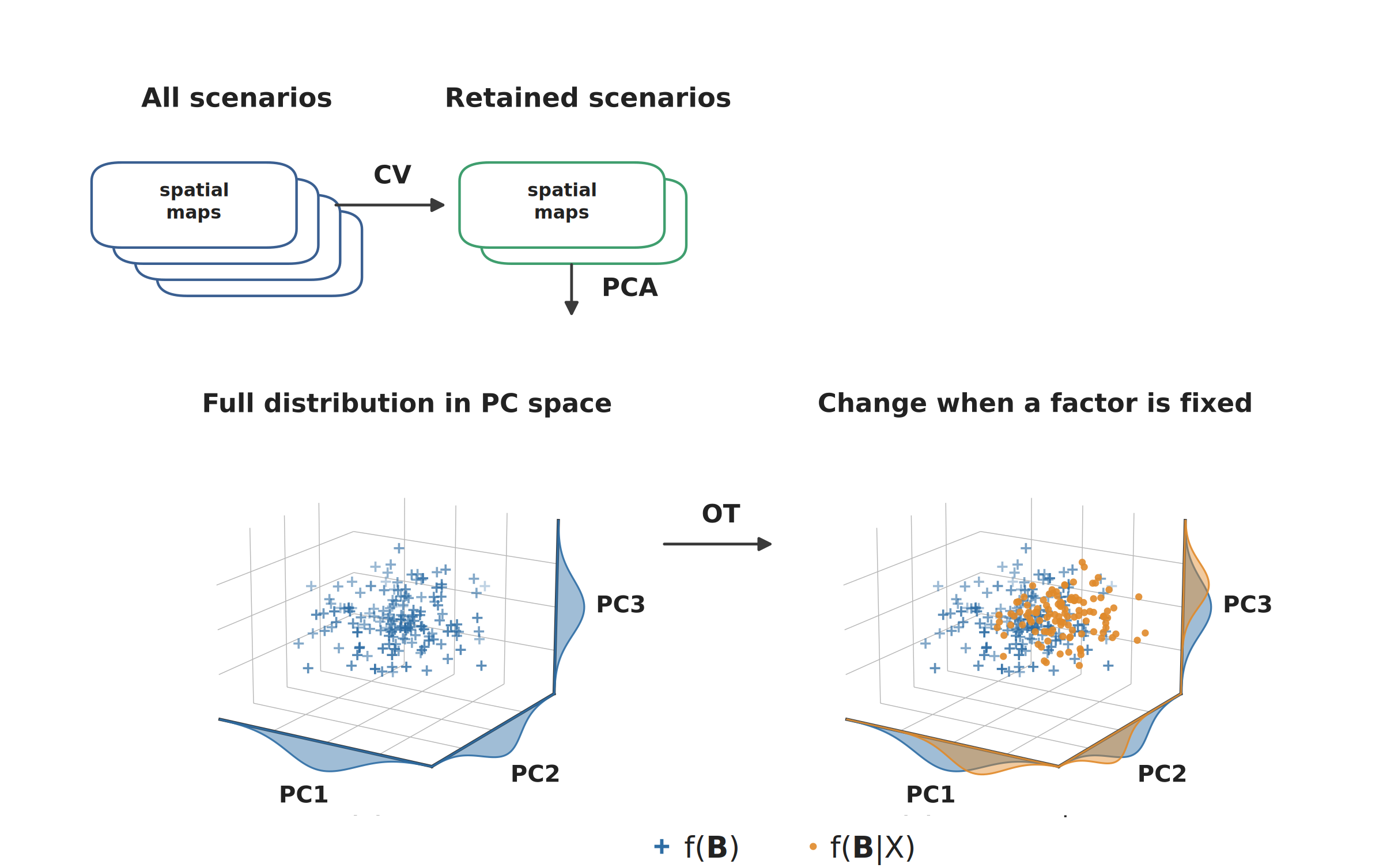}
    \caption{Schematic overview of the sensitivity analysis pipeline. 
    Scenarios are filtered by CV and projected into PC space by PCA.
For an uncertainty factor $X$, the
unconditional score distribution $f(\mathbf{B})$ (blue crosses) is compared by OT to the conditional distribution $f(\mathbf{B}\mid X\in
\mathrm{bin}_h)$ (orange points) for each bin of $X$. Averaging over bins gives the sensitivity index of $X$. }
    \label{fig:schematic_OT}
\end{figure}

Sensitivity analysis quantifies how the distribution of retained scenarios' scores $\mathbf{B}$ varies across modelling choices, using optimal transport (OT) sensitivity indices (\cite{Borgonovo2025}) to measure the shift induced by conditioning on a given factor, as summarised in Figure~\ref{fig:schematic_OT}.
For an uncertainty factor $X$, i.e.,  one of the $p$ columns of $\hat{\mathbf{U}}$ 
the parameter space is partitioned into $H$ bins. 
The OT index measures the expected discrepancy between the unconditional score distribution $f(\mathbf{B})$ and the conditional distribution obtained by fixing $X$ to each bin $h$, i.e., $f(\mathbf{B} \mid X \in \mathrm{bin}_h)$, averaged over the bins (\cite{Chiani2025}).
Formally,
\[
S_X
=
\frac{
\sum_{h=1}^{H}
w_h \,
R_{\varepsilon}
\!\left(
f(\mathbf{B}),
f(\mathbf{B} \mid X \in \mathrm{bin}_h)
\right) 
}
{2\displaystyle\sum_{r=1}^{K}\mathrm{Var}(B_{r})
} \in (0,1] \text{ with }
w_h=\frac{n_h}{\hat N},
\]
where  $n_h$ is the number of retained scenarios in bin $h$, $\hat N$ is the total number of retained scenarios,  $B_r$ denotes the vector of scores on the $r$-th component and $R_{\varepsilon}$ is the OT cost with entropic regularisation of strength $\varepsilon > 0$, chosen to be solved with the Sinkhorn algorithm (\cite{Cuturi2013}). 
The entropic Sinkhorn estimator, however, does not attain the lower bound:  so $S_X > 0$ even  when $X$ is independent of the output (\cite{Chiani2025}).
We therefore benchmark each index against an empirical irrelevance threshold, given by the index of a dummy factor drawn uniformly on $[0,1]$ and independent of the output by construction. 
This threshold is essential for the multivariate global index, whereas the per-PC indices use the exact one-dimensional quantile solver, for which the threshold captures only finite-sample noise.
Factors with indices at or below the threshold are treated as non-influential, whereas those above it contribute to variability in the resulting spatial predictions. 
Hence, a larger value of $S_X$ means that fixing $X$ shifts the score distribution  further from the unconditional distribution that carries all the uncertainty induced by all modelling choices jointly. 
 The factor with the largest index is thus the most influential.

Unlike variance-based Sobol' indices, the OT index is moment-independent: it
compares the full conditional and unconditional distributions, and captures changes in any feature of the distribution, not only in its variance. 
It is also defined for multivariate outputs, which is essential here because the score vector $\mathbf{B}$ is multivariate: the PCs are uncorrelated by construction but not necessarily independent (\cite{Jolliffe16}), so higher-order dependence between the scores would be invisible to a variance-based analysis but is retained by the OT comparison. For the same reasons the OT index is not additive and cannot be read as a percentage of explained uncertainty.

We report the analysis at two levels:
\begin{itemize}
    \item[1.] \textbf{Global SA:} provides an overview of the relative importance of each modelling factor but cannot reveal where on the map the sensitivity is concentrated. It rather answers the question: \textit{``Across all scenarios, how much does the overall spatial prediction vary?''}
    \item[2.] \textbf{Per-PC SA:}  analyses how each uncertain factor influences the distribution of scores on individual principal components, asking: \textit{``Which uncertainty sources drive the main spatial patterns of variation?''}
\end{itemize}

\section{Case study: groundwater sulphate concentration in the Paris Basin}
\label{sec:data}

We study groundwater sulphate concentrations (mg/L) in the Paris Basin, focusing on an upper Eocene (Bartonian) aquifer layer delineated by the entity 113AK of the BDLISA hydrogeological system.\footnote{https://bdlisa.eaufrance.fr/}
The sulphate in groundwater originates both from natural processes, such as the dissolution of minerals containing sulphates such as gypsum, and from anthropogenic activities, including agriculture, industry and mining.
Distinguishing between these natural and human-induced sources is a prerequisite for accurately assessing groundwater quality and identifying potential contamination (\cite{Sharma2020}). 
This aquifer studied here may contain high sulphates concentrations, as it is overlain by gypseous marls from the Lutetian age (upper Eocene). The presence of these mineral deposits in the form of beds or concretions is well known, and the region has several quarries where gypsum is exploited. 
 Gypsum is a common mineral in sedimentary environments and, being highly soluble, can reach elevated concentrations in natural groundwater.

The World Health Organization \cite{WHO2022} establishes no health-based guideline value, since available data do not identify a concentration likely to cause adverse effects. However, the presence of sulphate in drinking water can cause noticeable taste impairment, which varies with the nature of the associated cation: taste 
thresholds range from 250 mg/L for sodium sulphate to 1000 mg/L for calcium sulphate, and a laxative effect may occur in unaccustomed consumers at high concentrations, with notification of sources recommended above 500 mg/L.
In Europe, the Drinking Water Directive sets a quality limit of 250 mg/l.
\begin{figure}[h]
    \centering
    \vspace{-3mm}\includegraphics[width=0.7\linewidth, trim=0 25mm 0 25mm,clip]{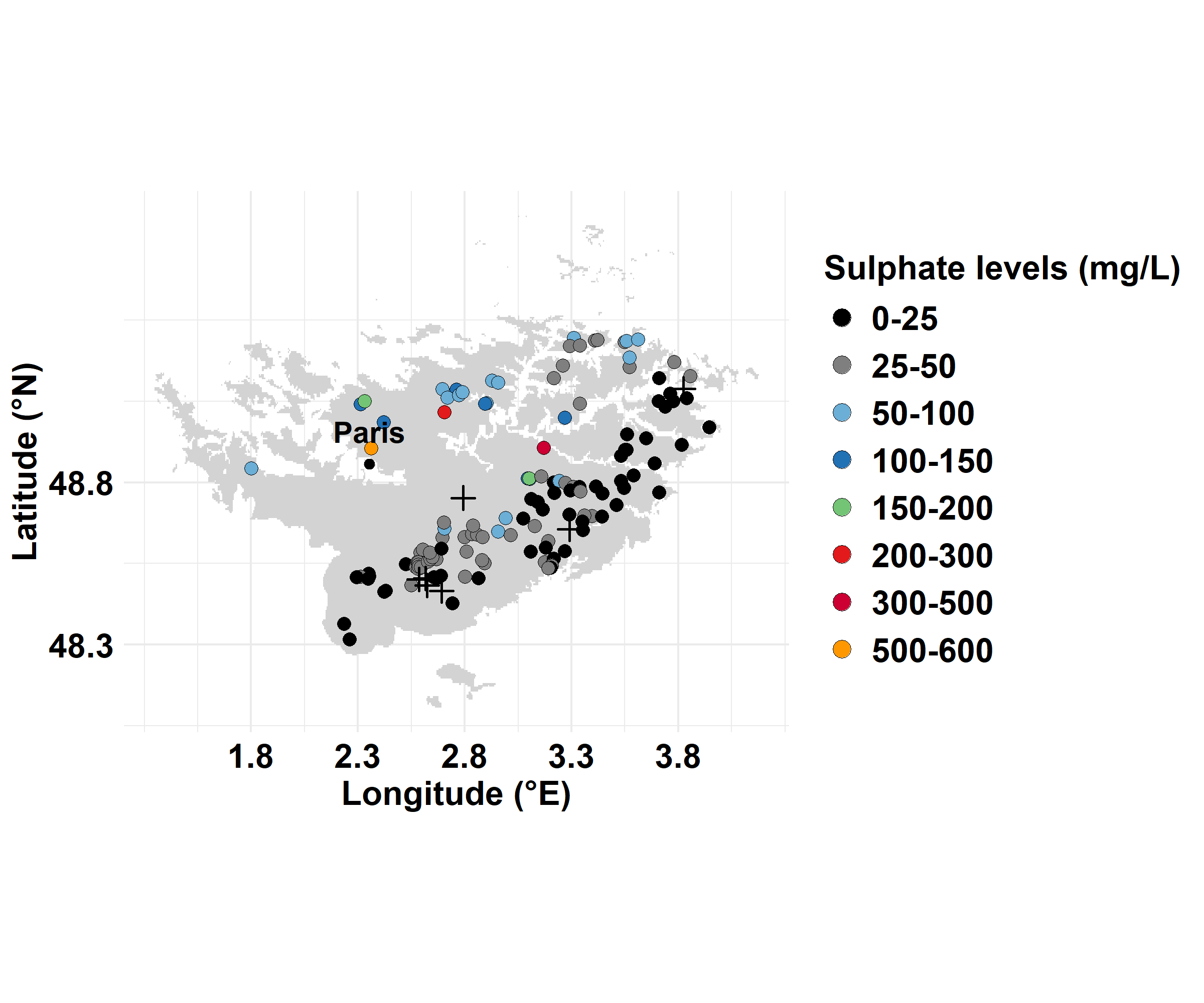}
    \caption{Spatial distribution of monitoring stations in aquifer 113AK, differentiated by sulphate concentration levels. Symbols distinguish between precisely measured concentrations (points) and censored observations (crosses).}
    \label{fig:data}
\end{figure}

Because stations are sampled repeatedly and irregularly through time, with records spanning April 1975 to May 2025, containing up to three measurements within a given month and up to fifteen within a given year, we reduce each station's record to a single representative value prior to spatial modelling by taking the temporal median. The median is robust to outliers and prevents frequently monitored stations from dominating the likelihood.
After aggregation, aquifer 113AK comprises 148 stations: 141 yield precise concentration values, while the remaining 7 are classified as censored measurements, see Figure~\ref{fig:data}. Based on the historical dataset, censoring thresholds decreased over time, from as high as 25 mg/L in some laboratories during the 1990s and early 2000s to 1 mg/L in recent analyses.

\subsection{The spatial model}
\begin{figure}[h]
    \centering
    \begin{subfigure}[b]{0.32\linewidth}
        \centering
        \includegraphics[width=\linewidth]{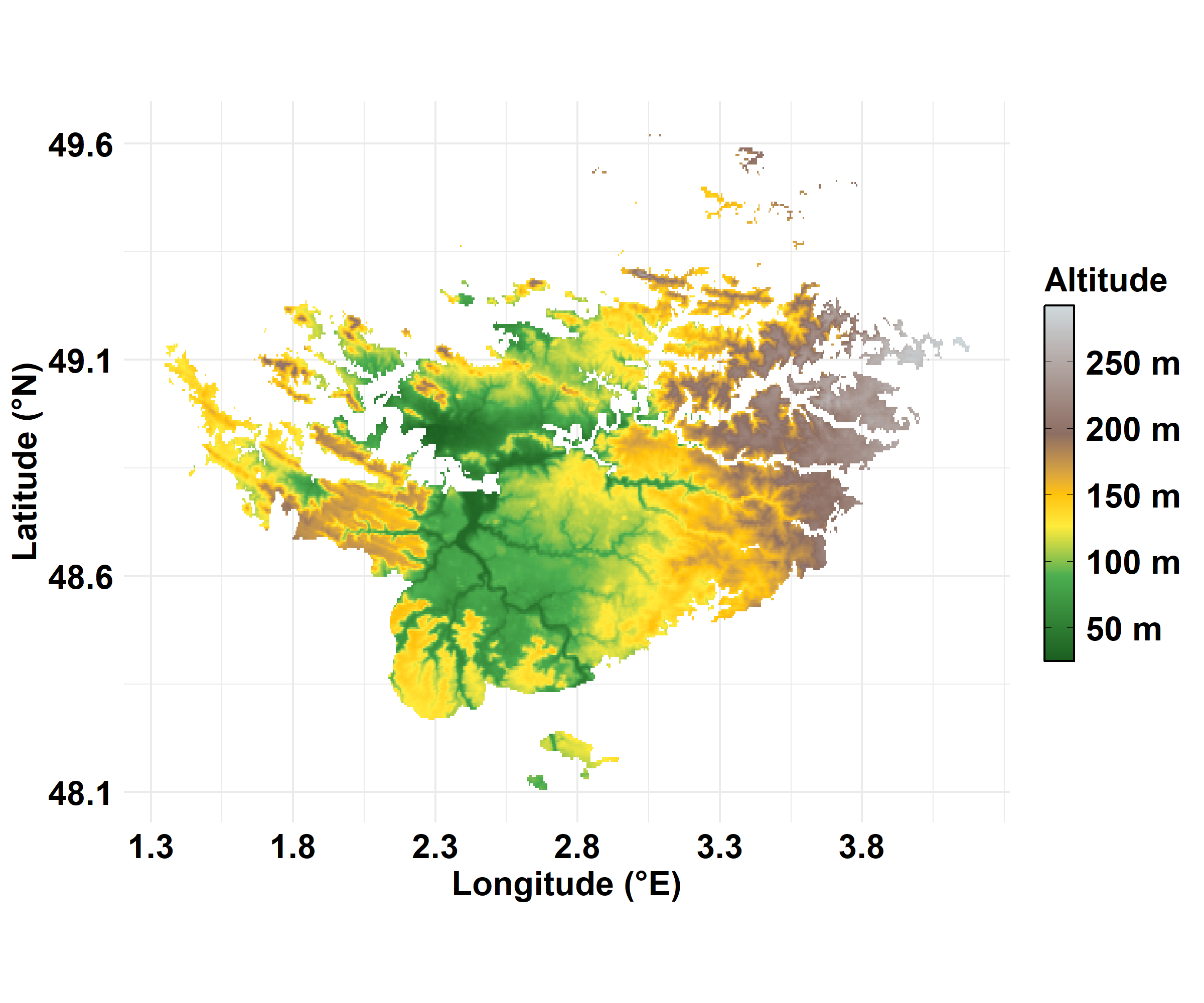}
        \label{fig:grid_1a}
    \end{subfigure}
    \hfill
    \begin{subfigure}[b]{0.32\linewidth}
        \centering
        \includegraphics[width=\linewidth]{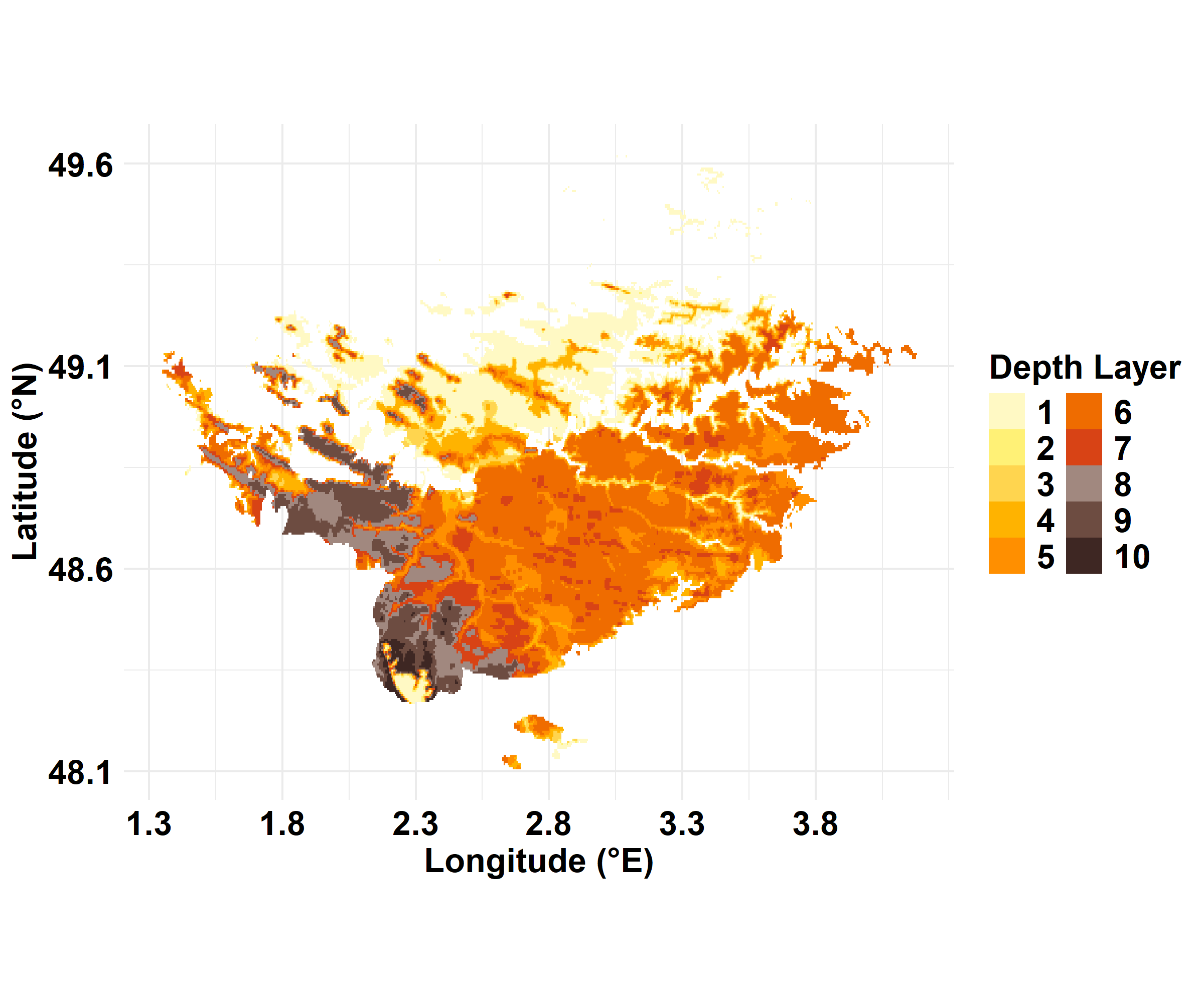}
        \label{fig:grid_1b}
    \end{subfigure}
    \hfill
    \begin{subfigure}[b]{0.32\linewidth}
        \centering
        \includegraphics[width=\linewidth]{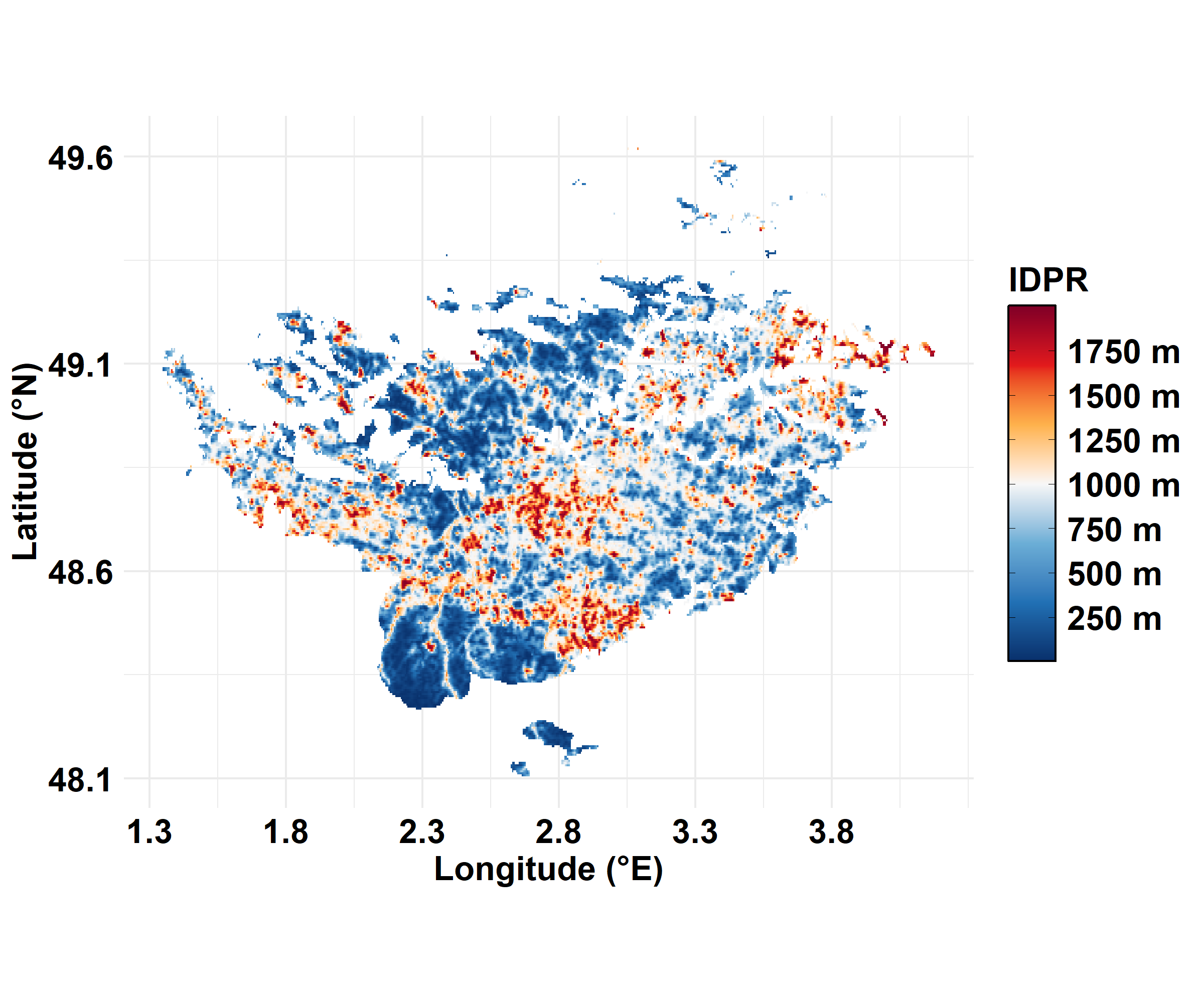}
        \label{fig:grid_1c}
    \end{subfigure}
    
    \vspace{-0.8cm}
    
    \begin{subfigure}[b]{0.32\linewidth}
        \centering
        \includegraphics[width=\linewidth]{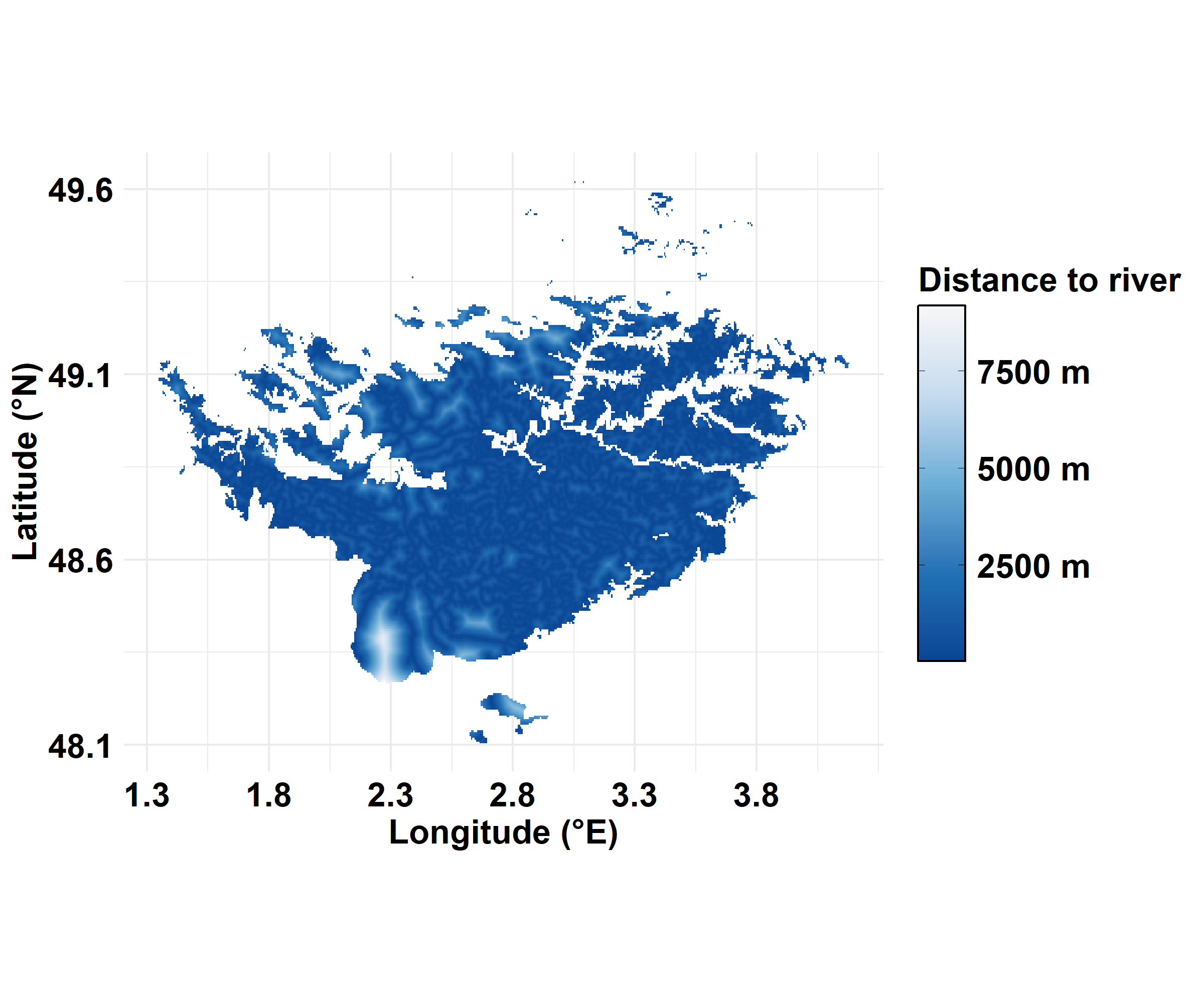}
        \label{fig:grid_2a}
    \end{subfigure}
    \hfill
    \begin{subfigure}[b]{0.32\linewidth}
        \centering
        \includegraphics[width=\linewidth]{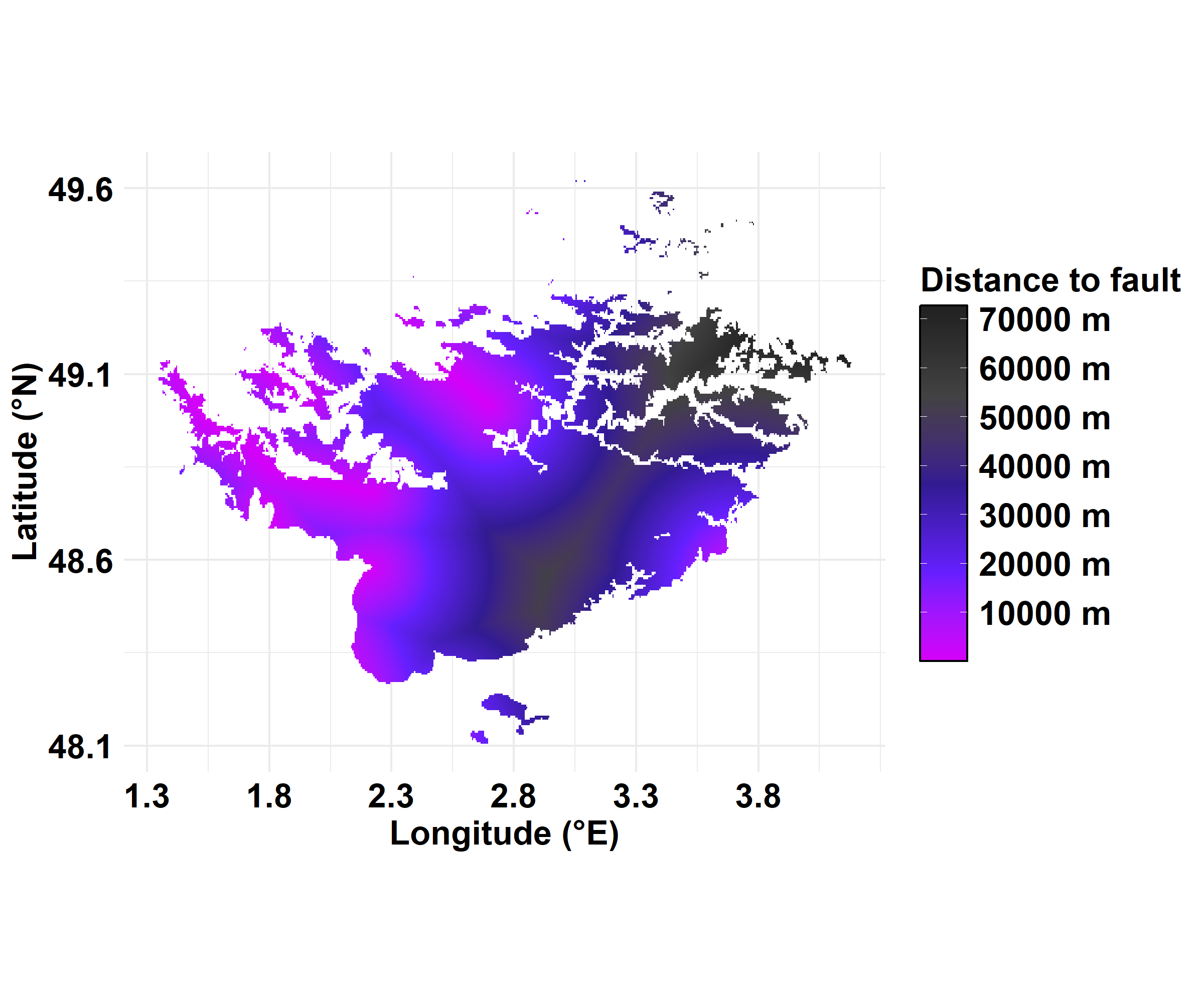}
        \label{fig:grid_2b}
    \end{subfigure}
    \hfill
    \begin{subfigure}[b]{0.32\linewidth}
        \centering
        \includegraphics[width=\linewidth]{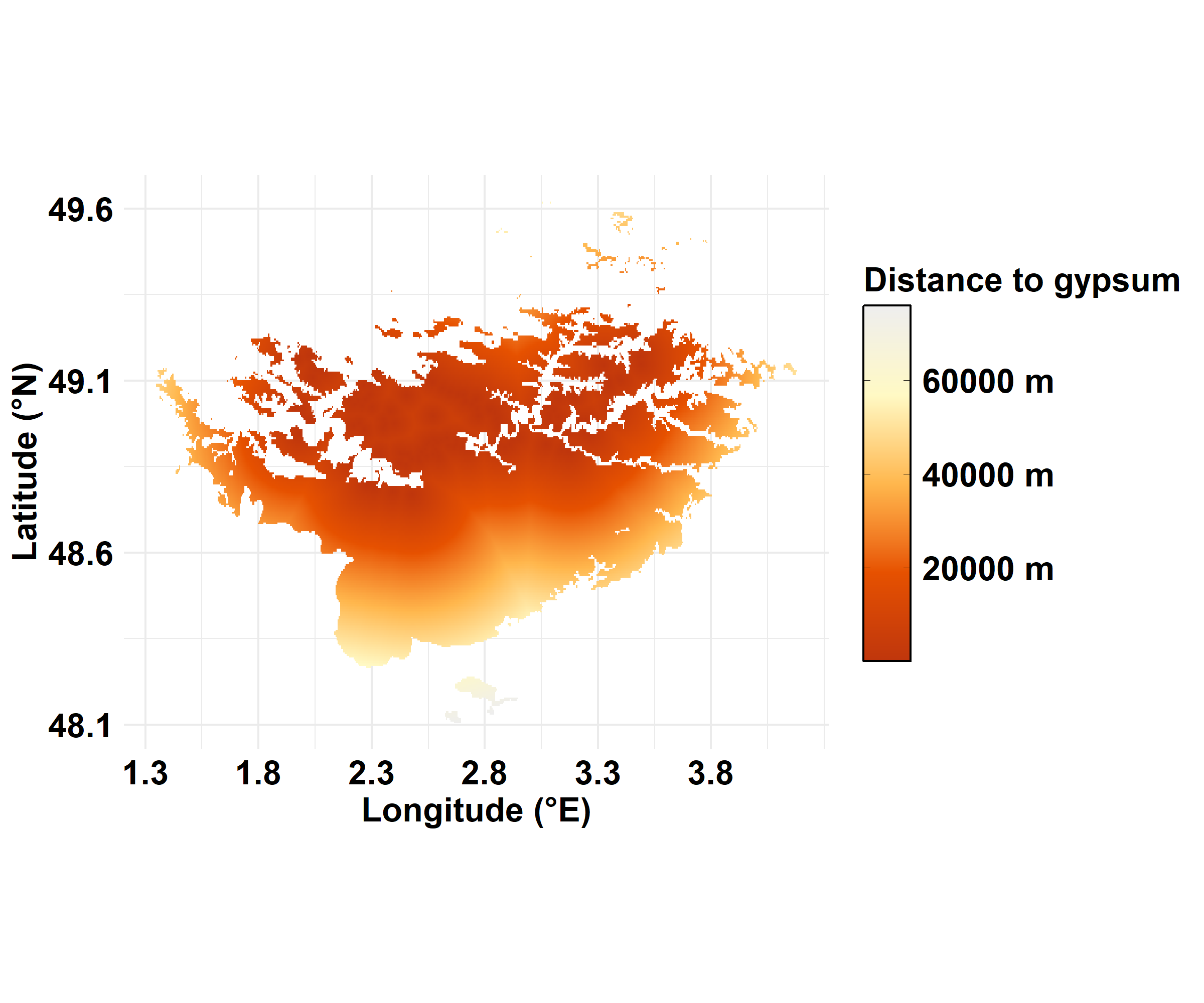}
        \label{fig:grid_2c}
    \end{subfigure}
\vspace{-8mm}
    \caption{Covariate maps from top left to right bottom: altitude (m), depth layers (categories), IDPR (dimensionless), distance to river, fault and gypsum sites (m).}
    \label{fig:main_six_grid}
\end{figure}

We adopt a Bayesian hierarchical model in which the covariates entering $f(\cdot, \beta)$ in Equation~\ref{eq:spatial_model} comprise altitude
(metres), depth layer (categorical), the network development and
persistence index (hereafter IDPR), and the Euclidean distances
(metres) to the nearest river, fault, and gypsum formation
(Figure~\ref{fig:main_six_grid}).
Altitude data are obtained from the French national open-data portal (\texttt{data.gouv.fr},  \cite{noauthor_modenumerique_nodate}). 
The depth layer variable describes the geological layer of the aquifer 113AK, corresponding to the Bartonian (Eocene) sands, limestones, and sandstones of the Paris Basin, as recorded in the BDLisa national hydrogeological reference database (\cite{brgm_base_2022,brugeron_brugeron_2018}).
A depth layer value of 1 indicates that the aquifer is the first geological layer beneath the ground surface. 
Higher values indicate that additional geological layers overlie aquifer 113AK, implying greater depth and, consequently, a lower potential of fluid infiltration from the surface. 
IDPR is a dimensionless value between 0 and 2000 that characterises  the favorability of infiltration (low IDPR$<1000$) versus runoff (high IDPR$>1000$), given topographic and hydrographic data (\cite{mardhel_description_2021}). 
River location data are obtained through the BD TOPAGE hydrographic database, accessed through the French national open-data portal (\texttt{data.gouv.fr}, \cite{noauthor_description_2020}).
Fault location are retrieved from the HIKE European Fault Database, as faults can be preferable fluid circulation paths.
Gypsum mineral is a known source of sulphate in groundwater, thus gypsum mining sites (gypsum quarries) are extracted from the 
 French national subsurface database (\cite{brgm_bss_2026}).
Distance to rivers, gypsum sites and faults are subsequently computed using the Euclidean distance tool from the ArcGis software (\cite{esri_arcgis_pro_2026}) and converted to raster data on a 100m long square grid following Lambert-93 (EPSG:2154) projection.
Covariates originally available in raster format (altitude, depth layer, and IDPR) are resampled onto the same 100m resolution Lambert-93 grid, which is defined to cover the full extent of aquifer 113AK. This ensures a common spatial support and resolution across all explanatory variables.

The spatial dependence in our model is represented by a random effect $Z(\mathbf{s})$ from  Equation~\ref{eq:spatial_model}, taken to be a Gaussian random field with a Matérn covariance function
\begin{equation*}
\mathrm{Cov}\big(Z(\mathbf{s}_i), Z(\mathbf{s}_j)\big)
=
\frac{\sigma^2}{2^{\nu-1}\Gamma(\nu)}
\big(\kappa \, \|\mathbf{s}_i - \mathbf{s}_j\|\big)^{\nu}
\, K_{\nu}\!\big(\kappa \, \|\mathbf{s}_i - \mathbf{s}_j\|\big),
\label{eq:matern}
\end{equation*}
where $\|\mathbf{s}_i - \mathbf{s}_j\|$ is the Euclidean distance between locations, $K_{\nu}$ is the modified Bessel function of the
second kind of order $\nu>0$, and $\Gamma$ is the gamma function.
 The field is thus governed by three parameters: the  smoothness $\nu$, the marginal standard deviation (SD) $\sigma$, and the
 scaling parameter $\kappa$, which is related to the more interpretable correlation range by $\rho = \sqrt{8\nu}/\kappa$. 
 We fix the smoothness through $\alpha = \nu + d/2 \in \{1.5,2\}$ with $d = 2$ throughout, which admits two natural members of the family: the exponential model ($\alpha = 1.5$, $\nu = 0.5$) and a smoother Matérn model ($\alpha = 2$, $\nu = 1$).
Rather than working with the dense covariance matrix of such a field, we use the stochastic partial differential equation (SPDE) approach of \cite{Lindgren2011}, whose discretisation on a mesh yields a Gaussian Markov random field with a sparse precision matrix.


The SPDE implementation requires priors on the field hyperparameters $\rho$ and $\sigma$. We use the penalised-complexity (PC) prior framework of \cite{Fuglstad2019}, which shrinks toward a base model of no spatial structure and is specified through interpretable tail-probability statements,
\begin{equation*}
    P(\rho < \rho_0) = \alpha_1
    \qquad \text{and} \qquad
    P(\sigma > \sigma_0) = \alpha_2 \qquad 0<\alpha_1, \alpha_2<1,
\end{equation*}
where the thresholds $\rho_0>0$ and $\sigma_0>0$ encode prior belief about the plausible correlation length and field amplitude informed by the empirical variogram. 
We set $\alpha_1 = \alpha_2 = 0.5$ so that the thresholds act as prior medians for the range and marginal SD. 

The SPDE solution is defined on a triangulation of the spatial domain: the continuous field is approximated by a piecewise-linear basis over the mesh, with the random effect evaluated at the mesh vertices and interpolated elsewhere. A mesh is therefore not optional, since it \emph{is} the discretisation of the latent field. Its resolution governs the quality of the approximation: too coarse, and fine-scale spatial variation is smoothed away. However, if it is too fine, computation becomes expensive without additional gains. The mesh must also be dense enough over the domain to resolve the spatial variability while remaining sparse in an outer extension that absorbs boundary effects at acceptable cost. 


\subsection{The inference method}


The SPDE approach yields a latent Gaussian model, the class for which the integrated nested Laplace approximation (INLA, \cite{Rue2009}) is designed. 
INLA is a Bayesian inference method that provides fast, deterministic approximations to the posterior marginals, which makes the present study feasible: fitting the $N$ models of the uncertainty matrix is tractable only because each individual fit avoids the convergence diagnostics and computational cost of Markov chain Monte Carlo. 
 Should a different inference method be adopted, e.g., kriging or Gaussian-process regression, the spatial model and its choices would be replaced by those of that method, while the remainder of the workflow is unchanged.


\section{Results}
\label{sec:results}

We now apply the six-stage workflow to the sulphate data of  Section~\ref{sec:data}, proceeding through the stages of Figure~\ref{fig:workflow}: We first  characterise the spatial dependence through the SPDE approach (Subsection~\ref{sec:diagnostics-results}), then we evaluate the candidate covariate structures and rank them to define the formula factor (Subsection~\ref{sec:formula-results}) and construct the uncertainty matrix (Subsections~\ref{sec:final_uncertainty}). 
The $N$ resulting scenarios are filtered by CV (Subsection~\ref{sec:cv-results}) and inferred on a high-resolution grid with INLA. The retained prediction maps are reduced by PCA (Subsection~\ref{sec:pca-results}) and finally analysed by OT sensitivity measures (Subsection~\ref{sec:sa-results}). 

\subsection{Choices on modelling spatial dependence}
\label{sec:diagnostics-results}
We now derive the levels of the three SPDE factors that constitute our sources of structural uncertainty: the covariance family, the PC-priors thresholds, and the mesh resolution.


To characterise the dependence structure, we estimate an empirical semivariogram and fit both exponential and Matérn covariance models 
parametrised by a nugget (measurement variance at the origin), a partial sill (the variance of the spatially structured process), and a scale parameter governing the decay of correlation.
Transferring the empirical fit to the SPDE parameters requires reconciling two conventions. First, the range: 
the variogram reports the scale parameter $a$ (\cite[Chapter~8]{Bivand2008}), whereas the SPDE prior is specified on the range $\rho$, the distance at which spatial correlation has decayed to approximately $0.13$. The two are related by $\rho = \sqrt{8\nu}\,a$ with $\kappa = \frac{1}{a}$. Second, the variance: the partial sill of the fitted variogram equals the marginal variance $\sigma^2$ of the SPDE field, so $\sigma = \sqrt{\text{partial sill}}$. Applying these conversions, the exponential fit ($\nu = 0.5$, $a = 7994$~m, partial sill $0.129$, nugget $0.004$) implies an effective range $\rho \approx 16000$~m and $\sigma \approx 0.36$, while the Matérn fit ($\nu = 1$, $a = 4563$~m, partial sill $0.108$, nugget $0.017$) implies $\rho \approx 12900$~m and $\sigma \approx 0.33$.

\begin{figure}[h]
    \centering
\includegraphics[width=0.8\linewidth]{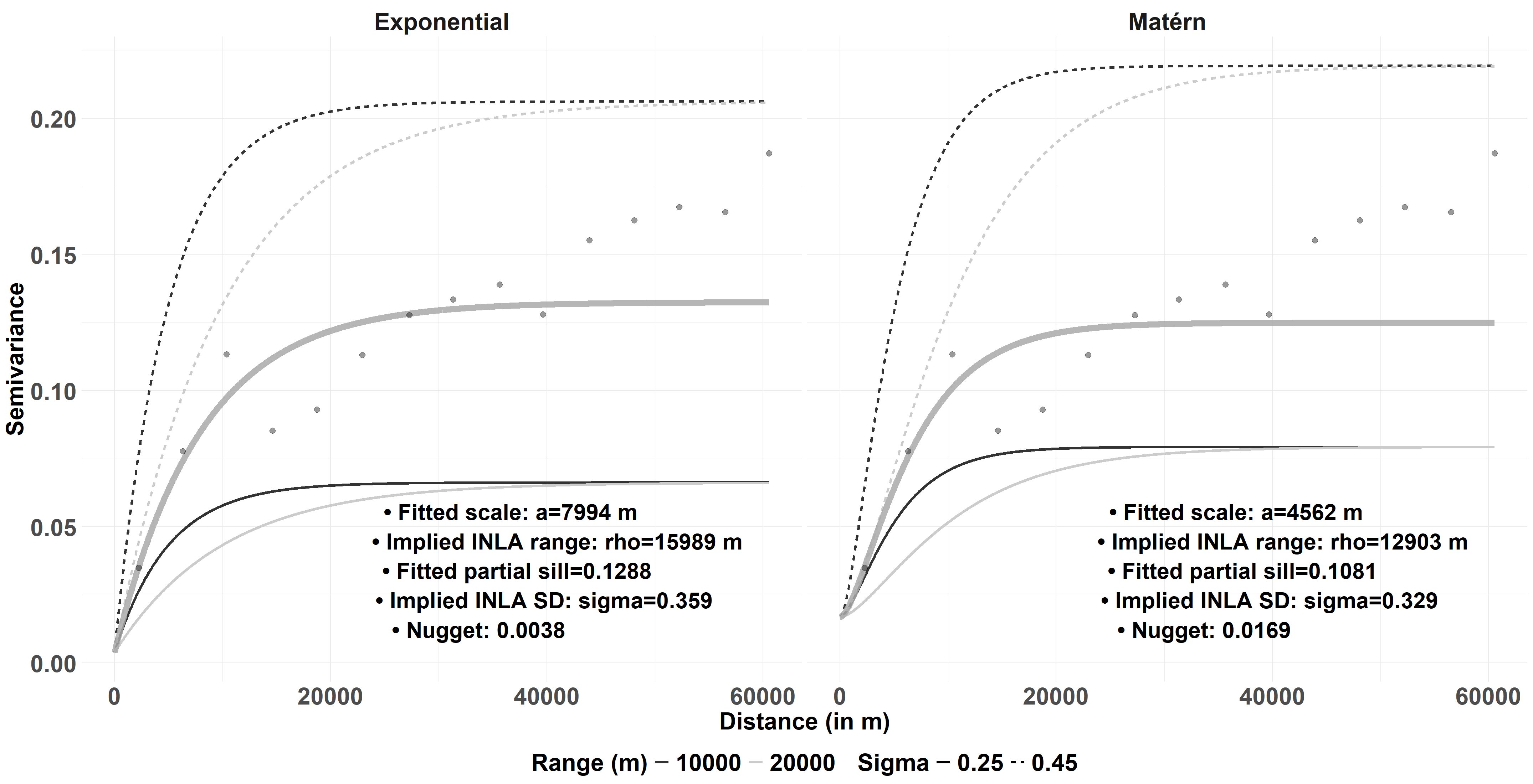}
\caption{Empirical fits and prior thresholds for the exponential ($\nu=0.5$) and Matérn ($\nu=1.0$) covariance models calculated across the aquifer study domain for the base model $\log_{10}(\text{sulphate concentration}) \sim 1$.}
\label{fig:ensemble_variogram}
\end{figure}

These fits inform the PC-prior thresholds. Figure~\ref{fig:ensemble_variogram} shows the empirical semivariogram (points) alongside the empirical fits (thick grey lines in the middle) and the upper and lower prior thresholds evaluated across different combinations of effective ranges and marginal SD. The fitted models place the effective range between roughly $[10000,\, 20000]$~m and the marginal standard deviation between $[0.25,\, 0.45]$. Guided by this, we sample the PC-prior thresholds from a $10\times10$ grid of $(\rho_0,\sigma_0)$ pairs uniformly over those intervals. 

\begin{figure}[h]
\vspace{-5mm}
    \centering
    \begin{subfigure}{0.45\linewidth}
        \centering
        \includegraphics[width=0.7\linewidth, trim=10mm 10mm 10mm 10mm,clip]{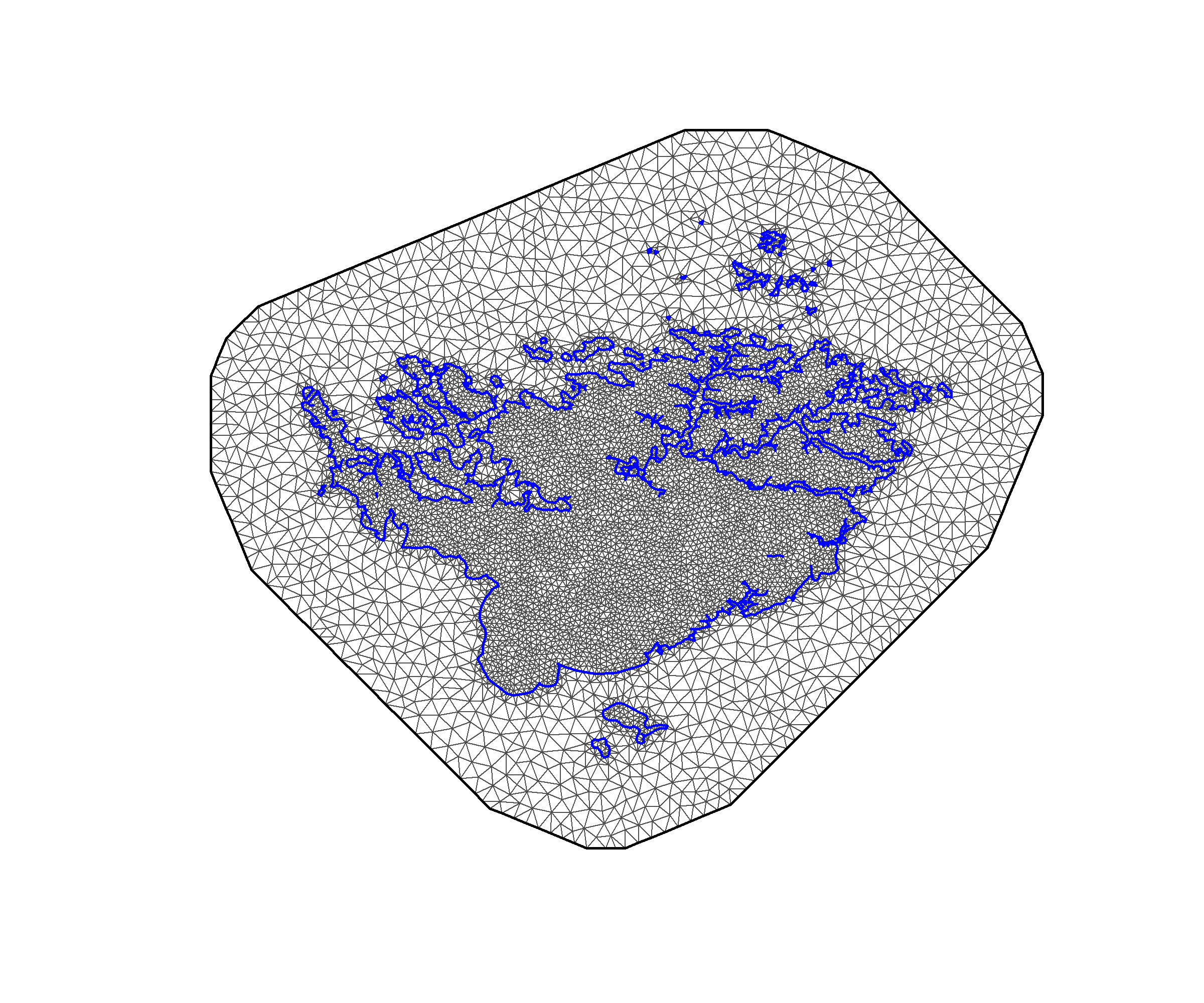}
        \caption{Dense mesh ($\sim$8700 nodes)}
        \label{fig:dense_mesh}
    \end{subfigure}
    \begin{subfigure}{0.45\linewidth}
        \centering
        \includegraphics[width=0.7\linewidth]{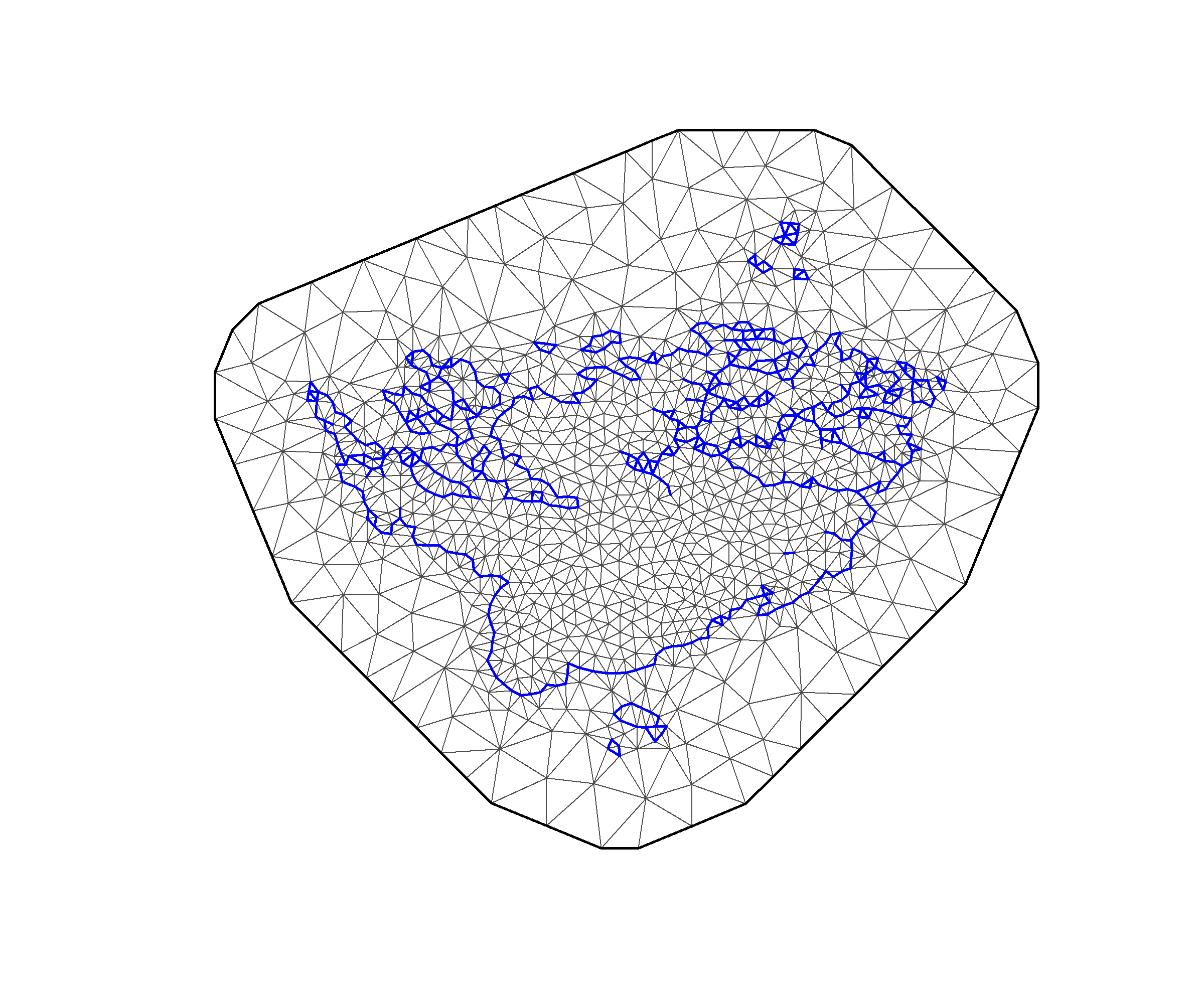}
        \caption{Coarse mesh ($\sim$1200 nodes)}
        \label{fig:coarse_mesh}
    \end{subfigure}
    \caption{Candidate meshes for aquifer 113AK. Inner domain (aquifer) uses smaller edges, while outer buffer uses larger edges to avoid boundary effects and reduce computation.}
    \label{fig:meshes}
\end{figure}

Rather than constructing meshes ad hoc, we generate the candidate triangulations from the mean of the empirical ranges, so that the discretisation is tied to the data-driven correlation length.
We parameterise each mesh by the maximum triangle edge length in the interior, which we vary across six levels spanning one third of to the full empirical range, and derive the remaining mesh parameters from it. 
The outer extension uses a coarser edge (1.5 times the interior edge) over a buffer of width one to two times the range, which displaces boundary effects away from the data, while the cutoff is set to one fifth of the interior edge to merge near-coincident vertices and avoid an ill-conditioned finite-element representation. This yields six meshes ranging from dense ($\sim 8700$ vertices, Figure~\ref{fig:dense_mesh}) to coarse ($\sim 1200$ vertices, Figure~\ref{fig:coarse_mesh}) 
assess sensitivity to spatial discretisation.

\subsection{Choices on modelling fixed effects}
\label{sec:formula-results}
Exploratory analysis indicates that altitude and distance to gypsum have non-linear relationships with the sulphate concentration variable. Therefore, these covariates are allowed to enter the model either as linear effects or as RW1  effects. Considering all possible combinations of inclusion or exclusion for the six covariates, and the alternative linear or RW1 specification for altitude and distance to gypsum, a total of $3 \times 3 \times 2^4 = 144$ candidate covariate structures are evaluated.
The spatial effect appearing in every model is a spatial random effect with  Matérn covariance function fitted via the SPDE approach with a fixed baseline specification (dense mesh from Figure~\ref{fig:dense_mesh}, and fitted prior thresholds from Figure~\ref{fig:ensemble_variogram}), so the DIC and WAIC rankings isolate the effect of covariate structure and likelihood.
Each formula is modelled through three candidate likelihoods: a Gamma, a log-Gaussian, and a Gaussian likelihood on $\log_{10}$-transformed concentration, and we rank the resulting 432 combinations by the DIC and WAIC criteria.
The Gaussian likelihood on $\log_{10}$-transformed concentration consistently dominated under both criteria, such that we report its top 10\% linear and non-linear covariate structures in Table~\ref{tab:formulas}. These 15 formulas, 7 linear and 8 non-linear,  define the formula factor of the uncertainty matrix.

\begin{table}[h]
\centering
        \scriptsize
        \begin{tabular}{lcc}
            \toprule
            \textbf{Top $10\%$ covariate combinations for $\log_{10}$(sulphate concentration)} & \textbf{DIC} & \textbf{WAIC} \\ 
            \midrule
            \multicolumn{3}{l}{\textit{Linear random effects}} \\
            \midrule
            altitude + dist. to gypsum + IDPR +  dist. to river + dist. to fault + spatial effect & 51.52 &  37.79\\
            altitude + dist. to gypsum + spatial effect & 51.73 & 38.77 \\
            altitude + dist. to gypsum + dist. to river + spatial effect & 52.16 & 38.86\\
            altitude + dist. to gypsum + IDPR + spatial effect & 53.17 & 39.94\\
            altitude + dist. to gypsum + dist. to river. + dist. to fault + spatial effect & 56.91& 43.65 \\
            altitude + dist. to gypsum + dist. to river. + dist. to fault + depth layer + spatial effect & 58.41 & 44.78 \\
            dist. to gypsum + IDPR + dist. to river + spatial effect & 60.42 & 45.94 \\
            \midrule
            \multicolumn{3}{l}{\textit{Non-linear random effects}} \\
            \midrule
            rw1(altitude) + rw1(dist. to gypsum) + IDPR + spatial effect & 42.87 & 26.87\\
            rw1(altitude) + rw1(dist. to gypsum) + spatial effect & 43.15 & 27.47 \\
            rw1(altitude) + rw1(dist. to gypsum) + IDPR + dist. to fault + spatial effect & 43.81 & 27.69 \\
            rw1(altitude) + rw1(dist. to gypsum) + dist. to fault + spatial effect & 45.95 & 30.09\\
            rw1(dist. to gypsum) + spatial effect & 46.32 & 30.27 \\
            rw1(altitude) + rw1(dist. to gypsum) + depth layer + spatial effect & 46.86 & 31.72\\
            rw1(altitude) + dist. to gypsum + IDPR + dist. to river + spatial effect & 46.91 & 31.70\\
            rw1(altitude) + dist. to gypsum + spatial effect & 47.33 & 32.91\\
            \bottomrule
            \vspace{0.5mm}
        \end{tabular}
  \caption{Top 10\% non-linear and linear covariate combinations  for $\log_{10}$-transformed data modelled with a Gaussian distribution and ranked by DIC and WAIC. rw1($\cdot$) represents a non-linear RW1 component.}
\label{tab:formulas}
    \end{table}

    \subsection{The uncertainty matrix}
    \label{sec:final_uncertainty}
The fitting stages of Subsections~\ref{sec:diagnostics-results}-\ref{sec:formula-results} fix the admissible levels of each factor:
\begin{itemize}
    \item \textbf{Covariance} (2 levels): exponential or Mat\'ern function both fitting the empirical
variogram comparably well.
    \item \textbf{PC-prior thresholds} (100 levels): a $10\times10$ grid of $(\rho_0,\sigma_0)$ pairs, sampled uniformly over intervals $[10000, 20000]$ and $[0.25,0.45]$ informed by the variogram fit, respectively.
    \item \textbf{Mesh resolution} (6 levels):  from dense to coarse triangulations.
    \item \textbf{Formula} (15 levels): the top 10\% linear and non-linear covariate structures explaining $\log_{10}$-transformed sulphate concentration data.
\end{itemize}
The full factorial of these $p = 4$ factors yields $2\times100\times6\times15 = N = 18000$ scenarios, so $\mathbf{U} \in \mathbb{R}^{18000 \times 4}$.

\subsection{Performance analysis and inference}
\label{sec:cv-results}
Our CV strategy partitions our data into validation groups based on the posterior correlation of the linear predictors, without requiring computationally prohibitive model refitting.
 The group size, i.e., the most informative data points to predict $y_i$  based on posterior correlations between the linear predictors,
 is set to its default. 
From the $N = 18000$ cross-validated scenarios, completed in $3.4$~h of computation, we retain for each evaluation metric the best-performing subset for the prediction stage. A scenario is retained under the aforementioned metrics from Subsection \ref{sec:CV}, if it falls in the most favourable decile of that metric across all scenarios: the lowest $10\%$ of $\log_{10}$-scale RMSE 
or the highest $10\%$ of $95\%$ coverage of the prediction interval.
This yields $\hat{N} = 2541$ retained scenarios, 
giving the reduced matrix $\hat{\mathbf{U}} \in \mathbb{R}^{2541 \times 4}$. 

\begin{figure}[h]
    \centering
    \vspace{-4mm}
    \includegraphics[width=0.5\linewidth]{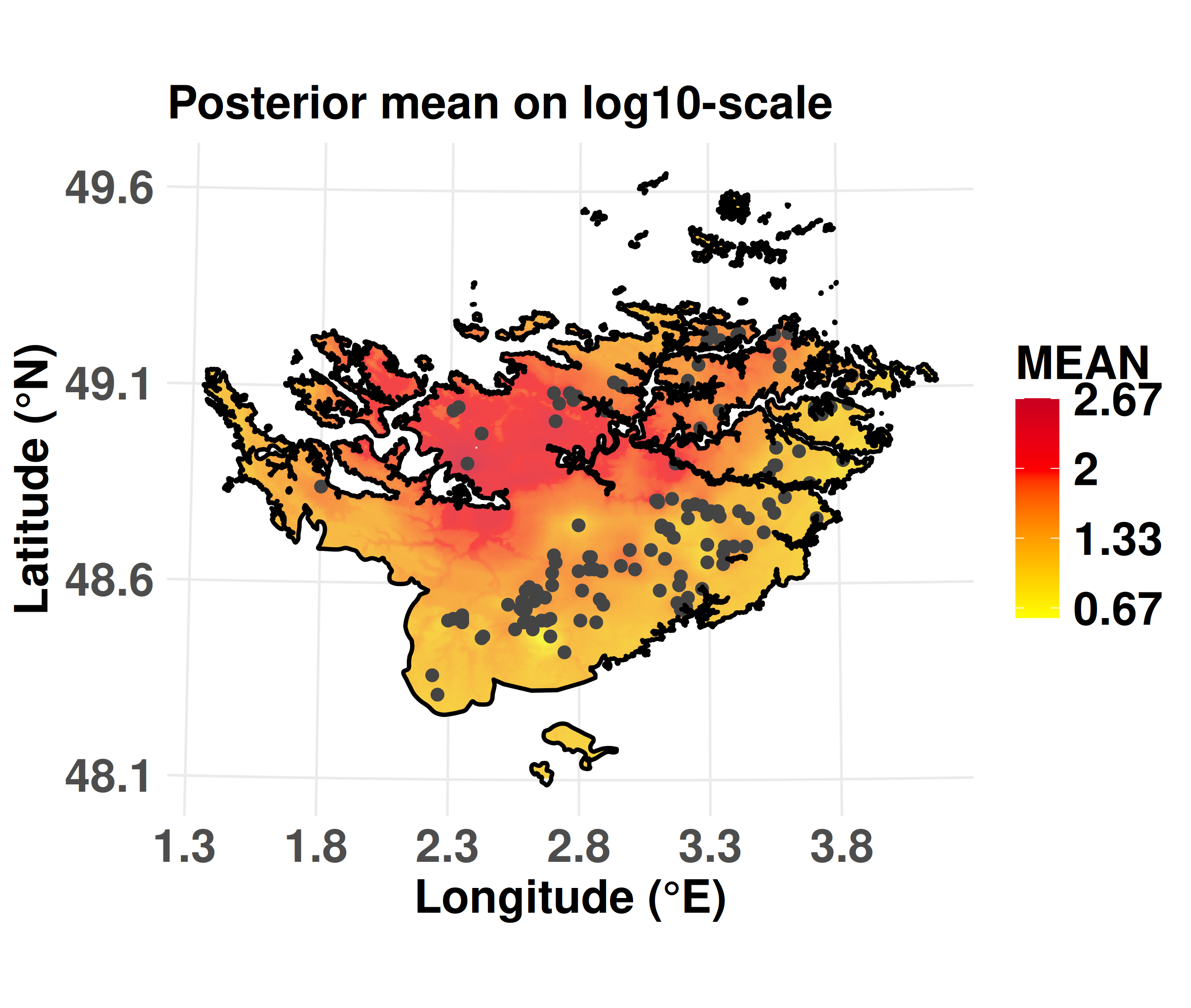}
    \caption{High-resolution prediction of the ensemble-average posterior mean of the $\log_{10}$-transformed data based on the filtering criteria regarding  RMSE and 95\% Coverage.}
    \label{fig:posterior_precitions}
\end{figure}
Predicting with INLA over the $G = 64500$ grid points for all $\hat{N}$ retained scenarios takes $1.5$~h and yields the ensemble mean matrix
$\mathbf{M} \in \mathbb{R}^{2541 \times 64500}$, which is passed to PCA.
Figure~\ref{fig:posterior_precitions} shows the ensemble-averaged posterior mean 
on the $\log_{10}$-scale, with monitoring stations (black points) overlaid.
The predicted posterior mean concentration is highest in the north-central part of the aquifer ($\log_{10}\!\approx 2.5$--$2.7$, i.e., $\approx\!300$--$500$~mg/L), decreasing south- and south-eastward to $0.5$--$1$ ($\approx\!3$--$10$~mg/L). 
This core coincides with the zone of minimum distance to gypsum (Figure~\ref{fig:main_six_grid}), consistent with gypsum dissolution as the dominant sulphate source.
 This area also corresponds to the section where the aquifer is shallower and is exposed to leaching by rainwater (IDPR $< 1000$), which causes the gypsum to dissolve. It is also anchored by an elevated local observation of $500$--$600$~mg/L, marked as orange in Figure~\ref{fig:data}.


\subsection{Dimension reduction}
\label{sec:pca-results}

PCA, applied to the centred and scaled ensemble as described in
Subsection~\ref{sec:pca-method}, identifies the dominant spatial patterns of variation across the retained scenarios.
The first three components explain $50.9\%$, $16.5\%$, and $12.4\%$ of the variance ($79.7\%$ cumulatively), while every further component explains less than $6\%$ individually. We therefore set the cumulative-variance threshold to $C = 70\%$.

\begin{figure}[h]
    \centering
\includegraphics[width=0.7\linewidth]{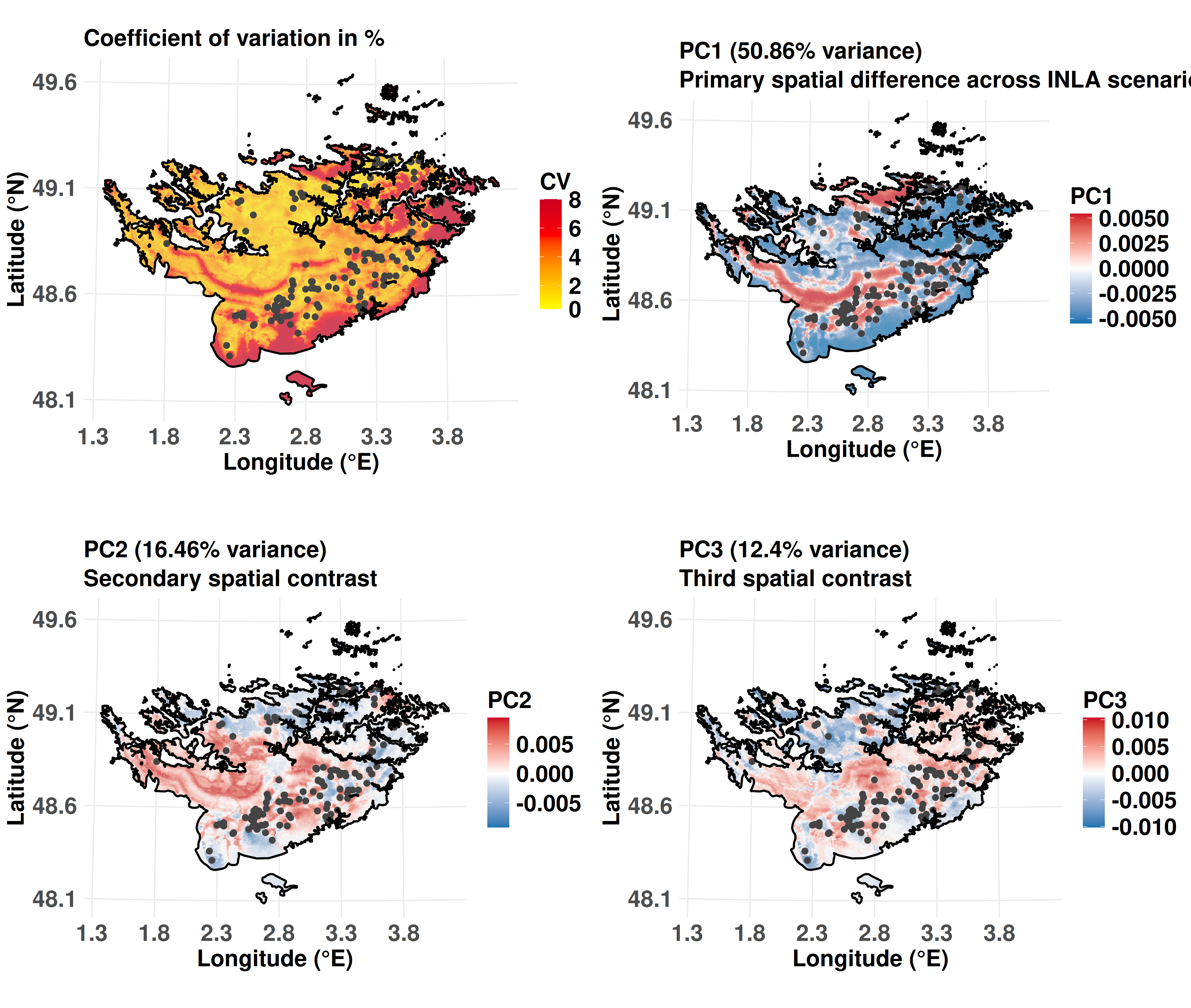}
    \caption{PCA on ensemble mean matrix $\mathbf{M}$  explaining $70\%$ of the variance  of the $\log_{10}$-transformed data based on the filtering criteria regarding  RMSE and 95\% Coverage.}
    \label{fig:pca_mean}
\end{figure}

Following \cite{Campbell2006}, who apply PCA to decompose functional model outputs for sensitivity analysis, we interpret the components as deviations about the ensemble mean map rather than as absolute concentrations.
In Figure~\ref{fig:pca_mean} the red and blue regions mark deviations in opposite directions across scenarios. 
The pixel-wise CoV is highest, $6$--$8\%$, along a central band and over the north-eastern and southern margins of the domain.

The leading component (PC1) is spatially coherent, contrasting a positive central band and northern rim against a predominantly negative interior, most strongly negative in the north-east.
Both extremes of PC1 coincide with the zones of elevated CoV, confirming that cross-scenario uncertainty is concentrated there.
Locations with high absolute PC1 loadings are therefore the most sensitive to changes in model configuration.
The second component (PC2) captures a finer contrast concentrated in the interior, showing that spatial uncertainty is not confined to the single large-scale pattern of PC1 but also has structure at a smaller scale.
 PC3 adds a further, weaker mode, with predominantly positive loadings across much of the central and northern domain. Together the three components show that uncertainty of the spatial patterns is structured at multiple scales rather than by one dominant gradient.
Because each grid point was standardised before decomposition, these maps locate where scenarios diverge \emph{relative} to the local spread. 

\begin{figure}[h]
    \centering
    \includegraphics[width=0.8\linewidth]{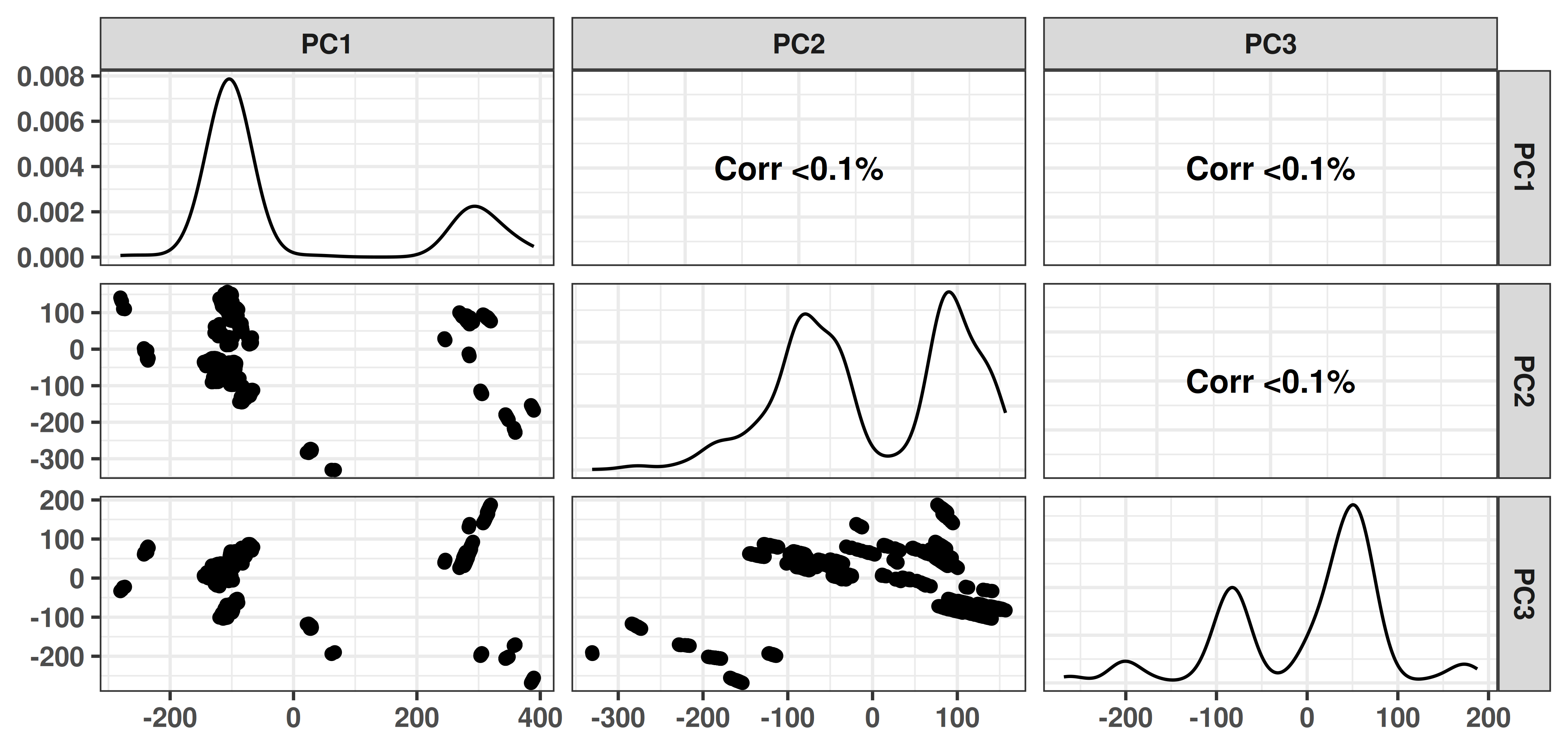}
    \caption{Diagonal panels show the marginal densities of the $\hat{N}=2541$ retained scenarios across the first three PCs, estimated by kernel density estimation. Lower panels show pairwise scatterplots. Upper panels report the pairwise correlations.}
    \label{fig:pc_pairs}
\end{figure}

Figure~\ref{fig:pc_pairs} shows the joint distribution of the retained scenarios across the three components. 
 Each marginal is multimodal, most markedly PC1, which separates the ensemble into two  groups with almost no scenarios in between. The ensemble therefore does not vary smoothly around a single central configuration: some modelling choices move entire groups of scenarios to a distinct region of the score space. 
 The pairwise scatterplots reveal that the components, although uncorrelated due to PCA (pairwise correlations below $0.1\%$), are far from independent: the scenarios form distinct clusters in every plane. The output is therefore a genuinely multivariate distribution whose dependence is non-linear.

A sensitivity measure based on variance alone, such as generalised Sobol' indices (\cite{LAMBONI2013, Gamboa2013}), would capture only
the spread of each component and the linear correlation between them, both uninformative here: the spread hides the bimodal splitting, and the correlations are zero by construction despite the evident dependence. This motivates the moment-independent and multivariate indices of the next Subsection, to quantify which modelling factors drive each PC and how strongly.


\subsection{Sensitivity analysis}
\label{sec:sa-results}
The OT indices depend on the number of bins $H$ into which each factor's range is divided. Following \cite{Chiani2025}, each bin should retain at least $100$
points, so with $\hat{N} = 2541$ retained scenarios we sweep $H$ in steps of two from $3$ to $25$ bins, the upper end leaving roughly $102$ points per bin. 
This sweep applies only to the continuous  factors, i.e., the prior thresholds on range and SD. The categorical factors (covariance, mesh resolution and formula) are instead binned by their own levels.

\begin{figure}[h]
    \centering
    \includegraphics[width=0.7\linewidth]{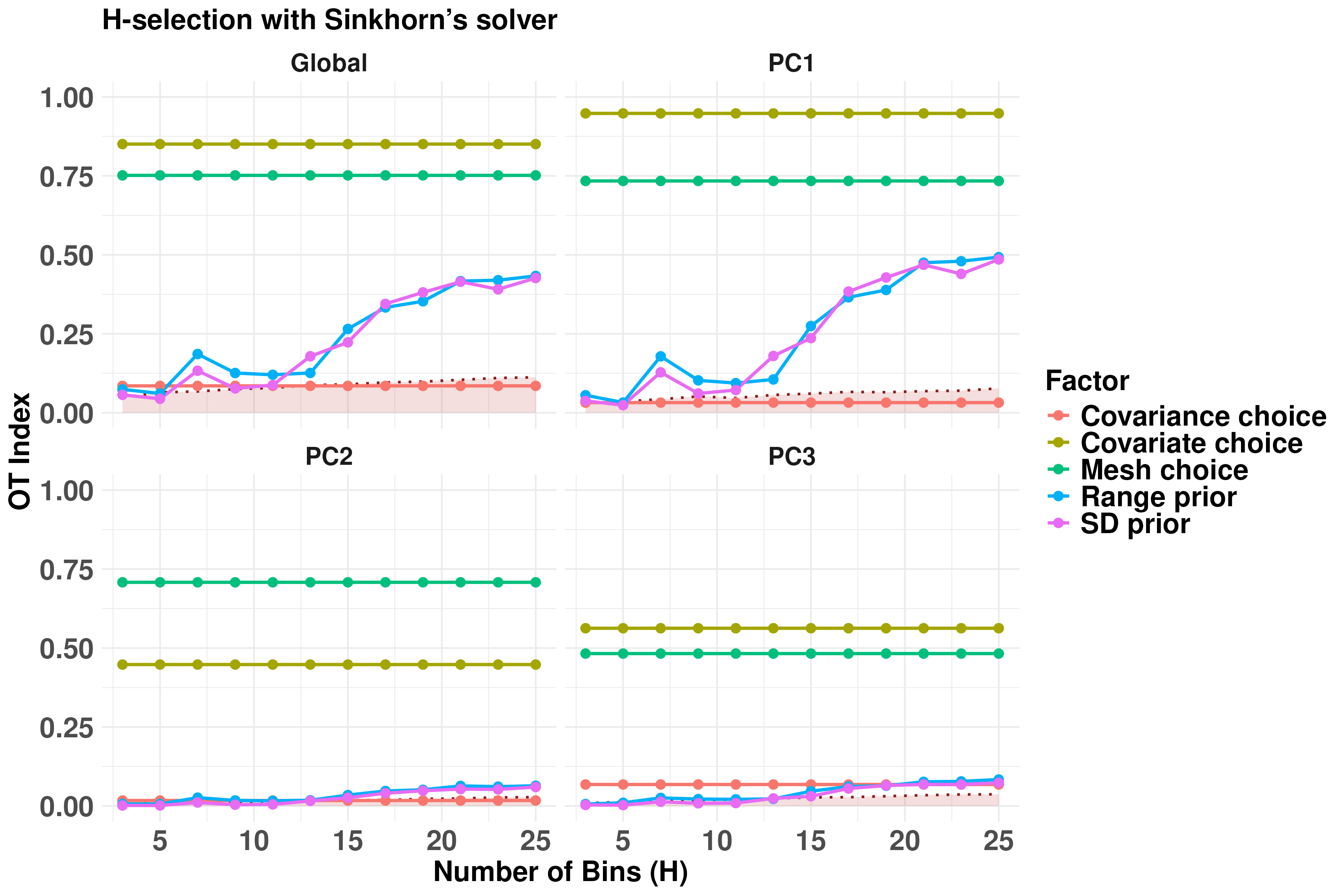}
    \caption{Selection of the number of bins $H$ for the OT-based SA. The chosen $H$ is the smallest value at which the relevant factors are clearly separated from the irrelevance threshold (red).}
    \label{fig:bin_selection}
\end{figure}

The Sinkhorn solver is run with regularisation $\varepsilon = 0.01$. 
 Factors with indices at or below the irrelevance threshold 
are treated as non-influential, while those above it contribute to the variability of the prediction maps. In Figure~\ref{fig:bin_selection} it becomes clear that the OT indices for the priors do not level out at a certain $H$, but rather increase gradually as $H$ increases.  
We select $H=20$ for the sensitivity analysis, corresponding to approximately $127$ scenarios per bin, as the smallest value yielding stable sensitivity conclusions while remaining well above the recommended minimum sample size per bin.

\begin{figure}[h]
    \centering
    \includegraphics[width=0.8\linewidth]{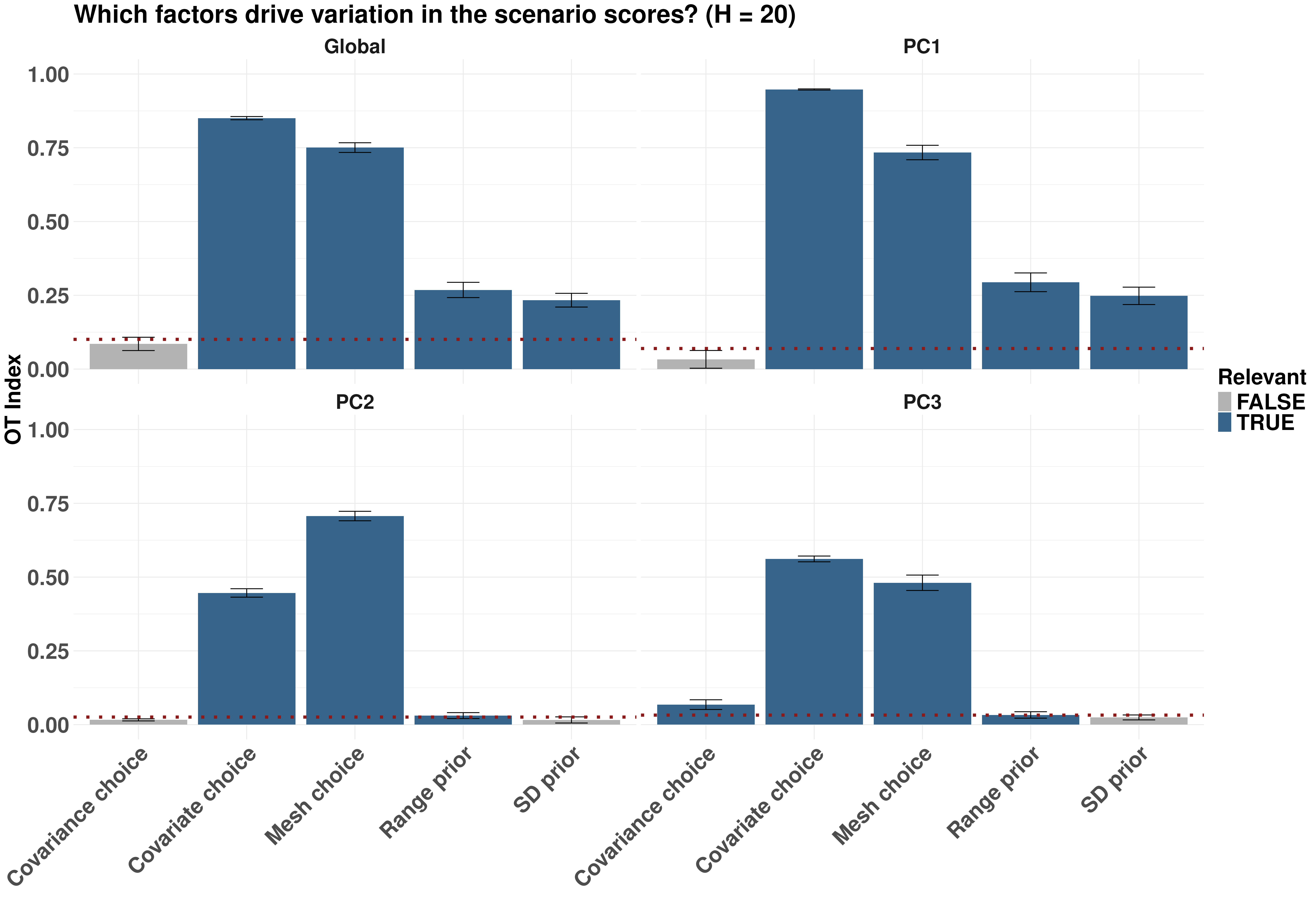}
    \caption{OT sensitivity indices based on filtering by RMSE on $\log_{10}$-scale and 95\% Coverage. Error bars show the 200-replicate bootstrap confidence intervals, and the dotted red line
the irrelevance threshold (95th percentile of the dummy draws).}
    \label{fig:SA_global}
\end{figure}

For the selected binning, and using $200$ bootstrap replicates to quantify estimation uncertainty, the global OT index identifies the  structural modelling decisions on covariate selection and mesh resolution as the dominant sources of variability, with $S_{\mathrm{covariate}} \approx 0.85$ ($95\%$ CI $[0.84, 0.86]$) and $S_{\mathrm{mesh}} \approx 0.75$ ($[0.73, 0.76]$), see Figure~\ref{fig:SA_global}. 
The thresholds on the spatial hyperparameter priors have a more moderate but significant effect, $S_{\mathrm{range}} \approx 0.27$ ($[0.24, 0.29]$) and $S_{\mathrm{SD}} \approx 0.23$ ($[0.21, 0.26]$). 
The covariance choice is the only factor  below the irrelevance threshold ($S_{\mathrm{covariance}} \approx 0.09$, CI $[0.06, 0.11]$ with  threshold $0.10$) and is therefore non-influential.
 The bootstrap confidence intervals are narrow throughout, indicating stable estimates.


The per-PC analysis refines this picture. 
On PC1 (irrelevance threshold $0.07$) the covariate choice dominates ($S_{\mathrm{covariate}} \approx 0.94$), followed by mesh resolution ($S_{\mathrm{mesh}}\approx 0.73$), with both thresholds on the priors significant but moderate ($\approx 0.29$ and $0.24$) and the covariance choice below threshold. 
PC2 is instead governed by the mesh resolution ($S_{\mathrm{mesh}} \approx 0.70$) ahead of the covariate choice ($S_{\mathrm{covariate}} \approx 0.44$). Here the choices on the covariance and SD prior threshold  play a negligible role, close to the irrelevance threshold of around $0.025$, while the threshold on the range prior only marginally exceed it.
The PC3  is mainly driven by the covariate choice  ($S_{\mathrm{covariate}}\approx 0.56$) ahead of the mesh resolution ($S_{\mathrm{mesh}}\approx 0.48$), while the remaining factors, excluding the threshold on SD prior, exceed the irrelevance threshold of $0.03$ only by small margins, thus having a rather weak contribution.

\section{Discussion}
\label{sec:discussion}
\subsection{Interpretation of results}

The sensitivity indices, the PC patterns, and the covariate plots together suggest a coherent physical reading. 
The covariate choice dominates PC1
($S_{\mathrm{covariate}} \approx 0.94$), whose positive central band coincides with the zone of elevated CoV in the centre of the aquifer. This band (Figure \ref{fig:pca_mean}) lies in the transition zone of the posterior mean between the gypsum-proximal northern part and the low-concentration south, where distance to gypsum decreases and altitude has either low or very high values. Both covariates enter  the candidate formulas either linearly or through RW1 effects, and it is in such transition zones that the two functional forms diverge most: a linear term imposes a constant gradient through the transition, whereas the RW1 term can follow plateaus. The concentration level in this band therefore depends strongly on which formulation a scenario uses, which is expressed as the dominant mode of PC1 and as elevated CoV.

The mesh resolution governs PC2 ($S_{\mathrm{mesh}} \approx 0.70$), whose finer-scale contrasts concentrate in the interior. This is consistent with the role of the triangulation: coarser meshes smooth between observations, while denser meshes allow more localised structure, so mesh resolution perturbs the prediction at the scale of inter-station distances rather than through a domain-wide gradient. The thresholds on the range and marginal SD prior have a moderate influence throughout ($S_{\mathrm{priors}} \approx 0.21$--$0.29$ globally), plausibly because the 148 observations are numerous enough to partially overwhelm the prior within the well-monitored belt, leaving prior sensitivity mostly to the data-sparse margins. The covariance family has globally a  negligible influence ($S_{\mathrm{covariance}} \approx 0.09$), in line with the exponential and Matérn fits describing the empirical variogram almost equally well  (Subsection~\ref{sec:diagnostics-results}), so that exchanging them alters the predicted surface only marginally.

\subsection{Sensitivity to the performance criterion}
\label{sec:robustness}
The retained ensemble depends on which CV metric defines ``best-performing'' and this choice is deliberately left to the analyst: different applications may prioritise point accuracy, calibration, a combination of the two or any other kind of metric. A useful property of the workflow is that it accommodates any such criterion without modification. Our primary analysis retains scenarios in the best decile of either $\log_{10}$-scale RMSE or $95\%$ prediction interval coverage. 
To illustrate this flexibility, we repeat the PCA and OT-based sensitivity analysis under two alternative criteria: the posterior mean ensemble selected solely on $95\%$ coverage, and the one selected solely on $\log_{10}$-scale RMSE.
The bin selection and PCA comparisons are reported in Appendix~\ref{app:X}. They show that the criterion governs not only the amount of retained variability  but also its spatial organisation. Here we examine how the influential factors change with the criterion.

\paragraph{OT-based sensitivity analysis}

\begin{figure}[h]
        \centering
    \includegraphics[width=0.7\linewidth]{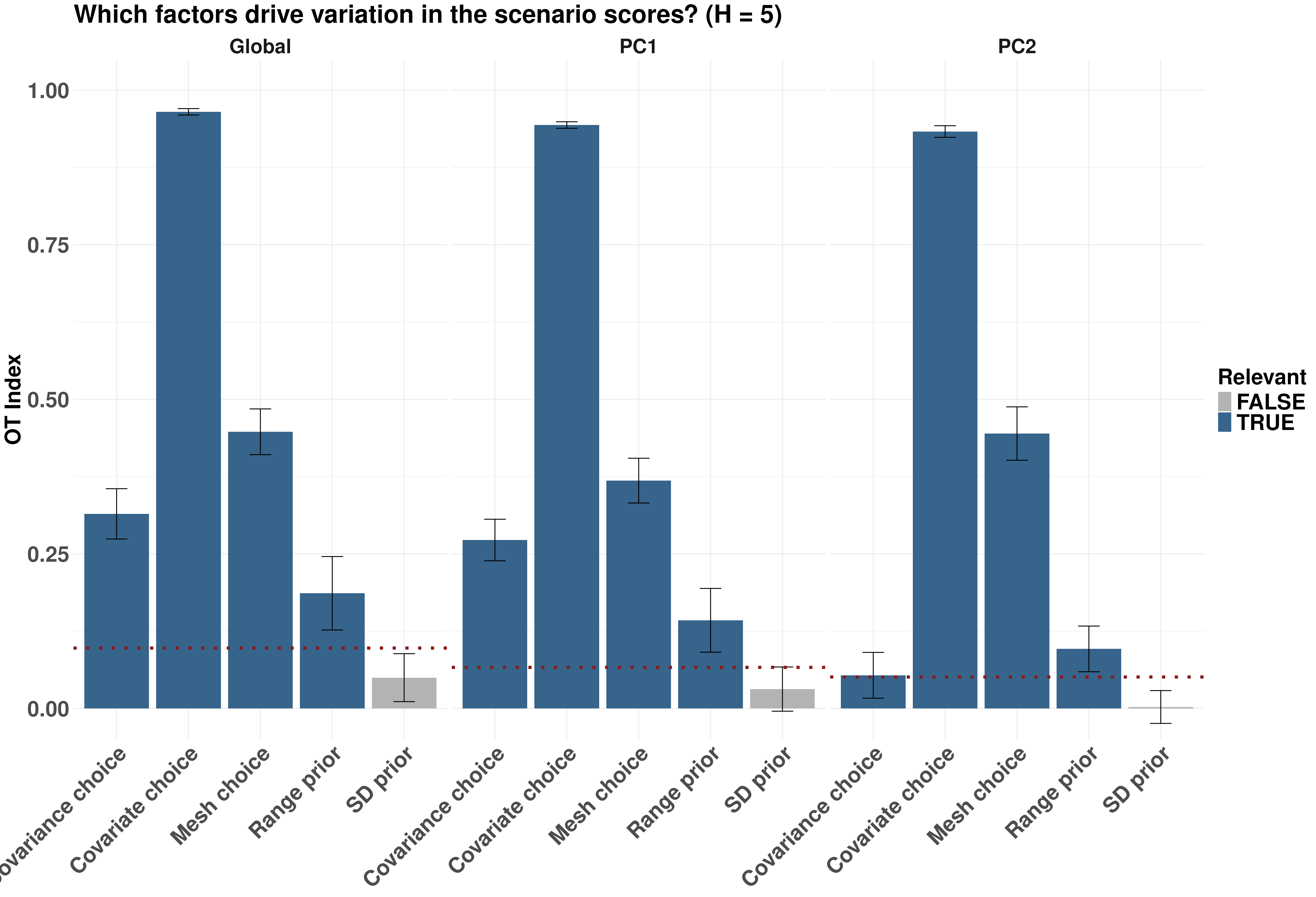}
    \caption{OT sensitivity indices based on filtering by 95\% coverage. Error bars show the 200-replicate bootstrap confidence intervals, and the dotted red line
the irrelevance threshold (95th percentile of the dummy draws).}        \label{fig:SA_cov}
    \end{figure}

\begin{figure}[h]

        \centering
        \includegraphics[width=0.7\linewidth]{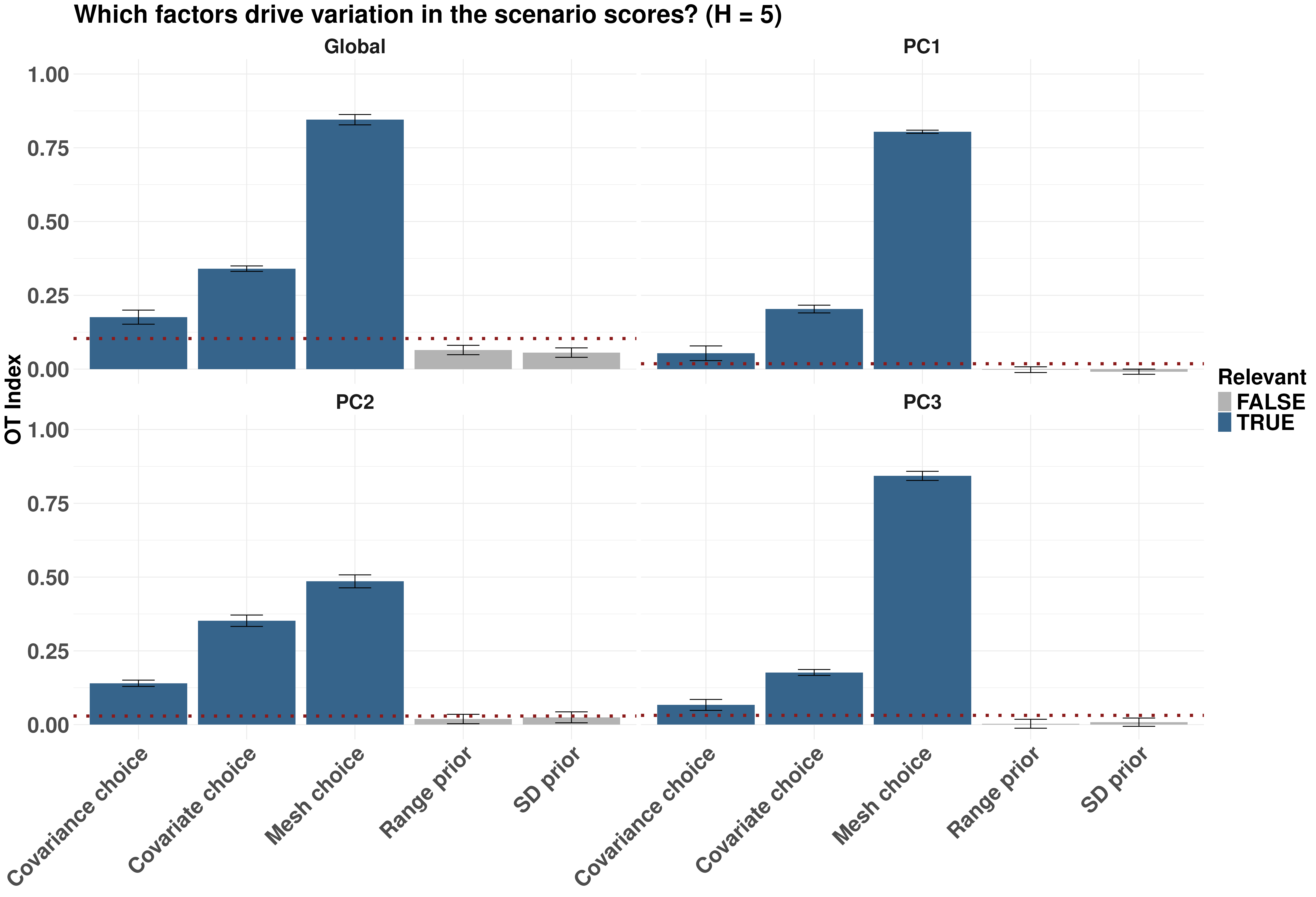}
        \caption{OT sensitivity indices based on filtering by RMSE on $\log_{10}$-scale. Error bars show the 200-replicate bootstrap confidence intervals, and the dotted red line the irrelevance threshold (95th percentile of the dummy draws).}
        \label{fig:SA_rmse}
\end{figure}

The OT indices in Figures~\ref{fig:SA_cov} and~\ref{fig:SA_rmse} confirm that the ranking of influential factors depends  on the criterion used to define the retained ensemble.
For the coverage-selected ensemble, the irrelevance thresholds are $0.10$, $0.07$, and $0.05$ for the global index and the first two PCs, respectively. 
The covariate choice dominates every level of
the analysis (global $S_{\mathrm{covariate}} \approx 0.96$, and $0.94$, $0.93$ on PC1 and PC2), followed by the mesh resolution ($\approx 0.45$ globally). 
The covariance family is relevant globally and on PC1 ($\approx 0.31$, $0.27$) but only marginally, and not significant on PC2. The threshold on the range prior has a modest effect ($\approx 0.09$--$0.19$), while the one for the SD prior stays below the irrelevance threshold throughout. 
Among models providing adequate predictive coverage, uncertainty is thus driven primarily by the choice of explanatory variables.

The RMSE-selected ensemble shows a markedly different structure. The irrelevance threshold is $0.10$ for the global index and $0.02$, $0.03$, and $0.03$  for PC1--PC3,
respectively. 
The mesh resolution is dominant throughout (global $S_{\mathrm{mesh}} \approx 0.85$ and $0.80$, $0.49$, $0.84$ on PC1--PC3), while the covariate choice is influential but secondary ($S_{\mathrm{covariate}} \approx 0.18$--$0.35$). 
The covariance family exceeds the threshold at every level (global $\approx 0.18$ and $0.05$, $0.14$, $0.07$ on PC1--PC3), though its contribution is small.
Both thresholds for the priors remain below the irrelevance threshold throughout and are non-influential.
For models selected on point accuracy, uncertainty is driven primarily by the mesh resolution.


\paragraph{Interpretation}
The two single-criterion ensembles isolate what each evaluation metric rewards, and the sensitivity indices shift accordingly. Retaining on RMSE alone keeps the scenarios that reproduce the observed concentrations most closely.
Within this ensemble the covariate-driven trend is largely pinned down, because a formulation that misplaces the regional gradient cannot rank in the best RMSE decile: the retained maps agree on the large-scale surface, and the CoV is correspondingly low and spatially uniform. The disagreement that remains lies in how the latent field interpolates \emph{between} the observations, which is governed by the triangulation, and the mesh resolution accordingly dominates every component.

Retaining on $95\%$ coverage alone reverses the picture. Calibration constrains the predictive intervals but places little restriction on the mean surface itself, so scenarios with substantially different regional trends survive the filter provided their uncertainty is honest. 
The retained ensemble is therefore the most diverse in its mean maps, visible in the higher CoV  and in a leading component that is a broad regional mode. The covariate choice dominates: the linear and RW1 formulations of altitude and distance to gypsum place the transition between the gypsum-proximal northern core and the low-concentration south differently (Figure~\ref{fig:main_six_grid}).
With no point accuracy criterion to discriminate between them, this trend disagreement carries through to the retained maps.
The covariance family and the threshold on the range prior also become relevant here, plausibly because both act on the field's correlation structure, and hence on the interval width that the coverage criterion evaluates, whereas under RMSE filtering their influence on the point prediction is minor.

Taken together, the criterion determines which modelling choice the retained ensemble remains sensitive to: point accuracy filtering fixes the trend and leaves mesh-scale interpolation as the residual uncertainty, while calibration filtering fixes the uncertainty behaviour and leaves the covariate-driven trend free. 
The combined analysis is consistent with this reading: in Figure~\ref{fig:SA_global} the covariate and mesh resolution jointly dominate, each inherited from one of the two criteria that define the combined ensemble.

\section{Conclusion}
\label{sec:conclusion}

We have introduced a workflow for attributing the uncertainty of a spatial
prediction map to the modelling choices behind it, combining a factorial design of choices, performance filtering, principal component analysis, and optimal transport global sensitivity analysis. 
The workflow is model-agnostic  and the added value is twofold.

 First, it is methodological: we adapt OT-based global sensitivity analysis to spatial outputs by coupling it with a  PC representation of the map ensemble.
 The resulting indices account for the multivariate, distributional nature of the output, an ensemble of maps rather than a scalar, which variance-based measures cannot address, and the associated PC maps localise where on the map each choice matters most.
Second, it quantifies for the first time the contribution of multiple modelling choices to spatial prediction maps derived in our case through the SPDE-INLA approach.
Choices  such as the covariance family or the covariate selection, are treated as controllable factors whose influence on the final map is measured rather than assumed. 
 Beyond these two points, the  sensitivity machinery is unchanged when our inference engine is exchanged for another, whether kriging, Gaussian process regression or a machine learning predictor,  the workflow transfers directly across spatial prediction problems and model classes.

Applied to groundwater sulphate concentrations in the Paris Basin, the analysis identifies the covariate choices in combination with their functional form and the mesh resolution as the dominant sources of uncertainty. The threshold choices for the priors are moderately important, and the covariance family is negligible, consistent with the exponential and Matérn models fitting the empirical variogram almost equally well. 
The dominance of these two factors reflects the two performance criteria: the mesh resolution emerges as dominant when scenarios are filtered on point accuracy (RMSE), and the
covariate selection when they are filtered on calibration ($95\%$ coverage).
The choice of performance criterion is thus itself part of the modelling uncertainty, not a neutral pre-processing step.
 Surfacing dependencies of this kind, which would remain hidden if a single  configuration and criterion are fixed, is precisely what the workflow is designed to reveal.


Several extensions are natural. A pixel-wise OT analysis would remove the reliance on the PC decomposition with a direct per-factor sensitivity map.
Its realisation would require efficient estimators for OT sensitivity maps. Such estimators could follow the fast basis-expansion methods recently developed for variance-based Sobol' maps (\cite{SAO2026}).
The framework could also accommodate different likelihood families or a spatio-temporal target, and, more broadly, serve as a template for uncertainty-aware spatial prediction wherever modelling choices are unavoidable and their consequences matter.
It could further integrate additional sources of error alongside the modelling choices, in  particular data imperfections arising from measurement error and from the spatial distribution of the samples (\cite{Wadoux2023}).



\subsection*{Code availability}
Variogram fitting is carried out with the \texttt{gstat} package (\cite{Pebesma2004}), and leave-group-out CV through R-INLA (\cite{Rue2009}).
Principal component analysis uses \texttt{prcomp} from the  \texttt{stats}  package (\cite{RTeam2011}), and the sensitivity analysis is performed with the \texttt{gsaot} package (\cite{Chiani2025}).
The data and code to the analysis are available on Github (\url{https://github.com/anrhouses/spatial_SA_with_OT}) and Zenodo (\url{https://zenodo.org/records/21725111}) (\cite{perchtold}). 

\section{Appendix}
\label{app:X}

\begin{figure}[h]
    \centering
   \includegraphics[width=0.7\linewidth]{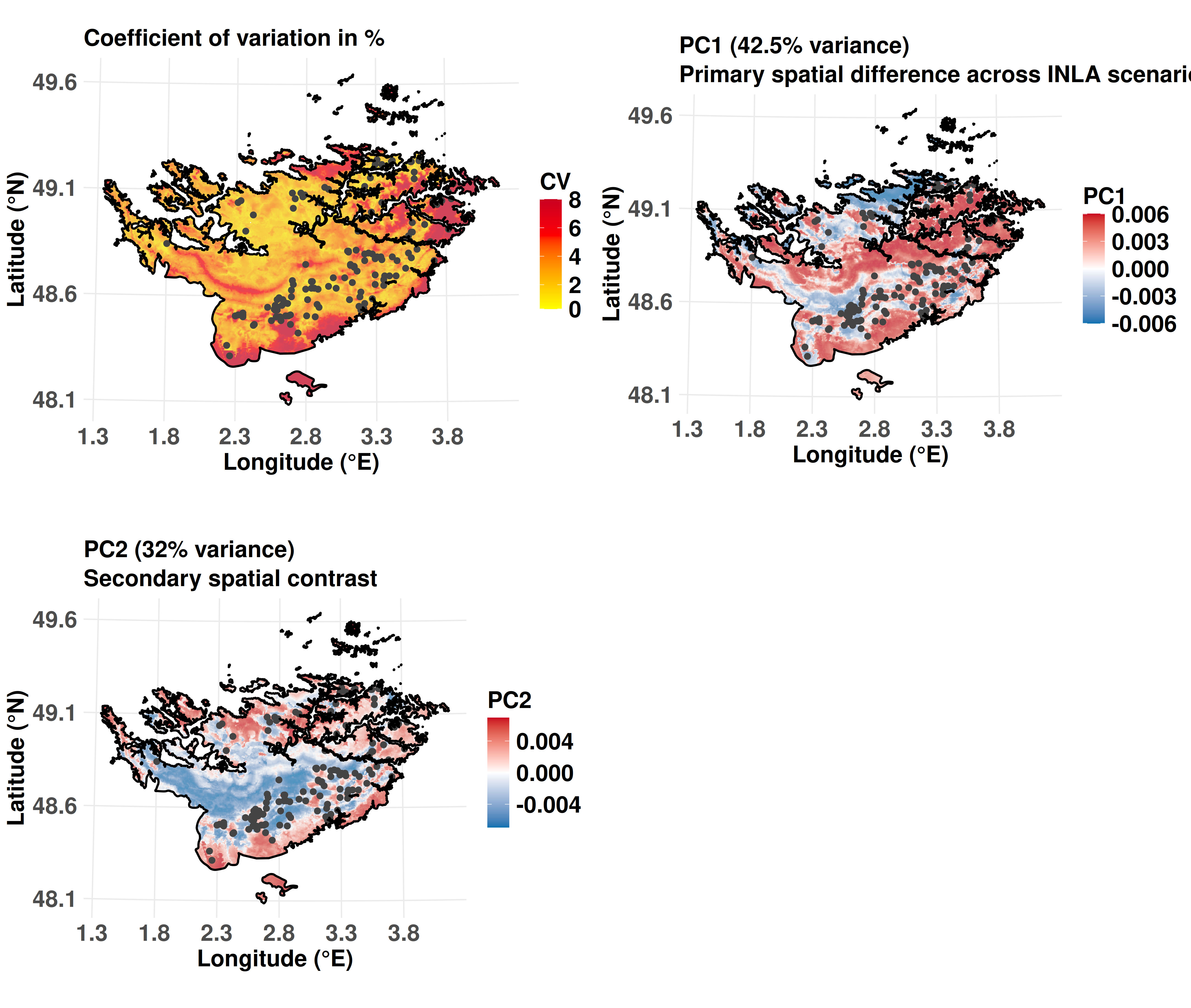}
   
    \caption{PCA on ensemble mean matrix $\mathbf{M}$  explaining $70\%$ of the variance  of the $\log_{10}$-transformed data based on the   95\%  coverage  individually.}
    \label{fig:pca_coverage}
\end{figure}

\begin{figure}[h]
    \centering
\includegraphics[width=0.7\linewidth]{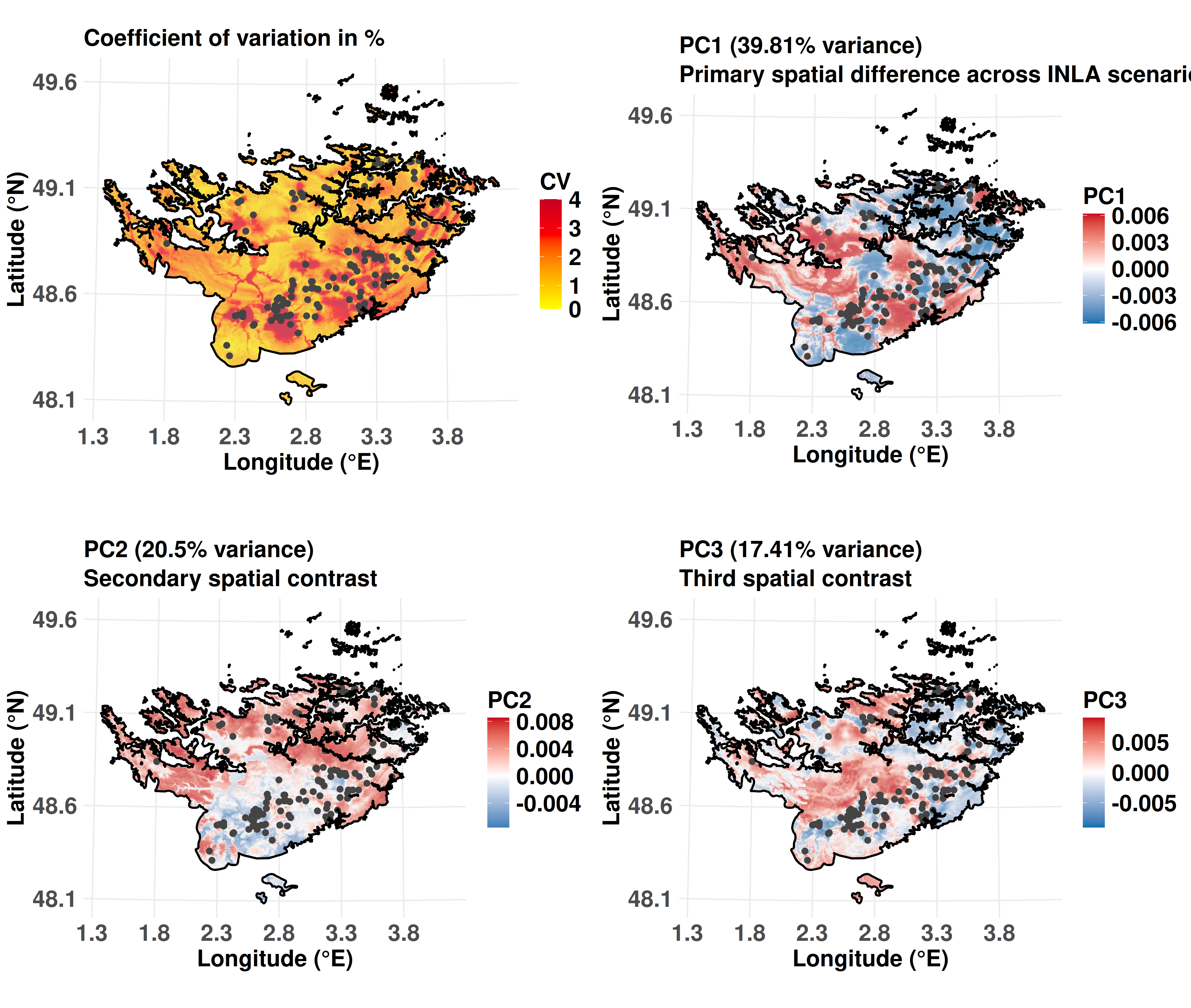}
 \caption{PCA on ensemble mean matrix $\mathbf{M}$  explaining $70\%$ of the variance  of the $\log_{10}$-transformed data based on RMSE individually.}
        \label{fig:pca_rmse}

\end{figure}

We report the PCA and bin selection for the
two single-criterion ensembles of Subsection~\ref{sec:robustness}, i.e., the
ensembles retained solely on $95\%$ coverage and solely on $\log_{10}$-scale
RMSE. These supplement the sensitivity results of the main text by showing how
the choice of filtering criterion shapes the retained uncertainty before the OT
analysis.
\paragraph{PCA} 
Figures~\ref{fig:pca_coverage} and~\ref{fig:pca_rmse} show that the choice of filtering criterion substantially affects both the magnitude and spatial structure of the retained uncertainty. 
The coverage-selected ensemble exhibits larger and more extensive variability, with CoV reaching approximately $8\%$, against more localised values around $4\%$ for the RMSE-selected ensemble.

The decompositions also differ.
For the coverage-based selection, two principal
components already explain $74.5\%$ of the variance (PC1 $42.5\%$, PC2 $32\%$), so scenario-to-scenario variability reduces to two dominant spatial gradients. The RMSE-based selection instead requires a third component (PC1 $39.81\%$, PC2 $20.5\%$, PC3 $17.41\%$), indicating a wider range of alternative spatial representations among scenarios retained on predictive accuracy alone.
The PC1 patterns differ in scale and location.
For the coverage-selected ensemble, PC1 highlights positive loadings primarily along the central, north-eastern and southern margin of the aquifer.
For the RMSE-selected ensemble, PC1 concentrates mainly in the central part of the aquifer, near the area containing the highest observed sulphate concentration and the  cluster of monitoring stations. The most substantial negative loadings occur predominantly towards the northern and eastern margins.
This suggests that models selected solely for predictive accuracy disagree most strongly on the spatial extent and magnitude of sulphate concentration.
The second principal component of the coverage-selected ensemble reveals a marked north-south contrast and accounts for $32\%$ variance compared to $20.5\%$ for the RMSE ensemble.

\begin{figure}[h]
        \centering
        \includegraphics[width=0.7\linewidth]{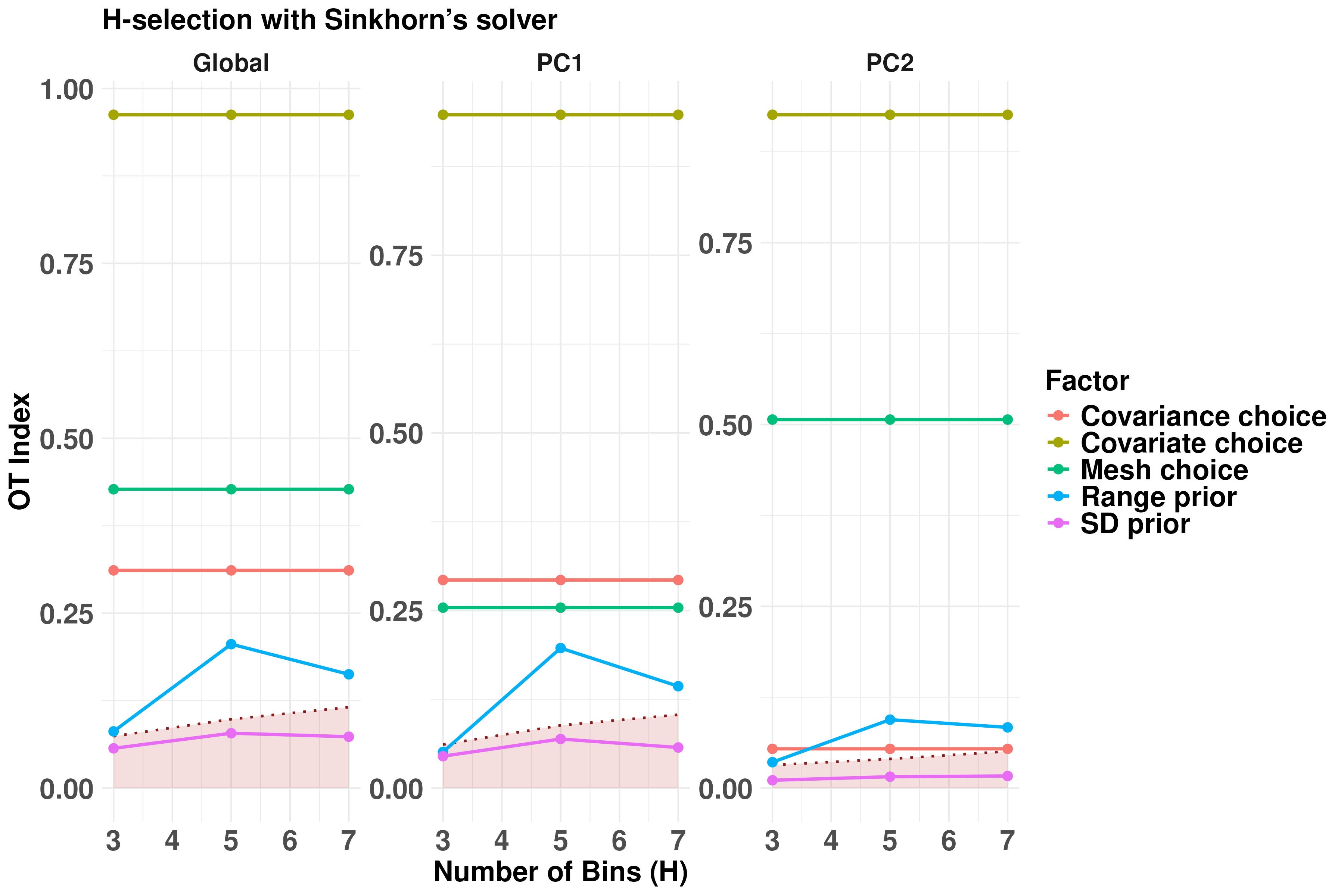}
    \caption{Selection of the number of bins $H$, based on 95\% coverage only, for the OT-based SA. The chosen $H$ is the smallest value at which the relevant factors are clearly separated from the irrelevance threshold (red).}        \label{fig:stage1_cov}
    \end{figure}

\begin{figure}[h]
    \centering  
        \includegraphics[width=0.7\linewidth]{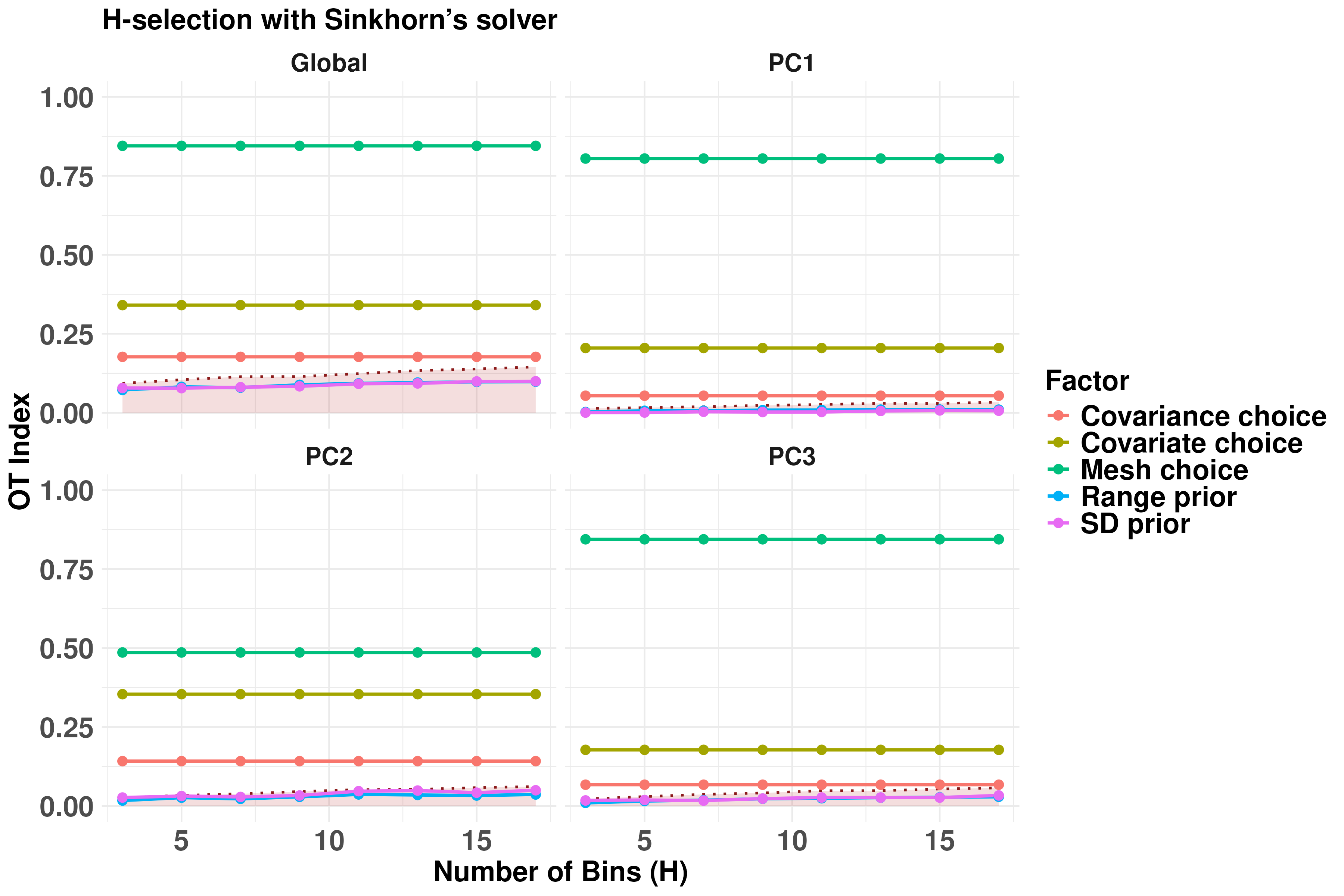}
        \caption{Selection of the number of bins $H$, based on RMSE only, for the OT-based SA. The chosen $H$ is the smallest value at which the relevant factors are clearly separated from the irrelevance threshold (red).}
    \label{fig:stage1_rmse}
\end{figure}

\paragraph{Bin selection}
Following the procedure of the main analysis, the number of bins $H$ is selected across a range of admissible values (Figures~\ref{fig:stage1_cov} and~\ref{fig:stage1_rmse}).
For the smaller coverage-selected ensemble ($\hat{N}=741$) we consider $H \in \{3,\ldots,5\}$ and for the RMSE-selected ensemble ($\hat{N}=1800$) we sweep through $H \in \{3,\ldots,17\}$,  respecting the requirement of at least $100$ scenarios per bin. 
In both ensembles, the factor ranking is stable across $H$, with only minor changes in index
magnitude, and $H=5$ is retained for the final analysis.

\subsection*{Authors contribution}
Corinna Perchtold did the investigation, formal analysis, methodology, software, validation, visualization and writing.
Jérémy Rohmer did the funding acquisition, conceptualization, methodology, supervision, proof-reading and editing.
Augustin Thomas and Julie Lions did the  data curation. 

\subsection*{Conflict of interests}
The authors declare that they have no conflict of interest. 


\subsection*{Acknowledgements}
This research was supported by the French National Research Agency within the HOUSES project (grant ANR-22-CE56-0006). 
The authors acknowledge BRGM for providing SAPHIR computing and storage resources and Boulahya Faiza for her support on the setup.
During the preparation of this work, the author used Claude in order to improve the language and grammar of this manuscript. After using this tool/service, the author reviewed and edited the content as needed and takes full responsibility for the content of the publication.


	\bibliographystyle{plain}
\bibliography{lib_tex.bib}

\end{document}